\documentstyle[psfig]{mn}

\newif\ifAMStwofonts

\ifoldfss
  \newcommand{\rmn}[1] {{\rm #1}}

  \ifCUPmtlplainloaded \else
    \NewTextAlphabet{textbfit} {cmbxti10} {}
    \NewTextAlphabet{textbfss} {cmssbx10} {}
    \NewMathAlphabet{mathbfit} {cmbxti10} {} 
    \NewMathAlphabet{mathbfss} {cmssbx10} {} 
  \fi
  \ifAMStwofonts
    \ifCUPmtlplainloaded \else
      \NewSymbolFont{upmath} {eurm10}
      \NewSymbolFont{AMSa} {msam10}
      \NewMathSymbol{\upi}     {0}{upmath}{19}
      \NewMathSymbol{\umu}     {0}{upmath}{16}
      \NewMathSymbol{\upartial}{0}{upmath}{40}
      \NewMathSymbol{\leqslant}{3}{AMSa}{36}
      \NewMathSymbol{\geqslant}{3}{AMSa}{3E}

       \let\le=\leqslant
       
    \fi
  \fi
\fi 

\ifnfssone
  \newmathalphabet{\mathit}
  \addtoversion{normal}{\mathit}{cmr}{m}{it}
  \addtoversion{bold}{\mathit}{cmr}{bx}{it}
  \newcommand{\rmn}[1] {\mathrm{#1}}

  \def\textbfit{\protect\txtbfit}
  
  \long\def\txtbfit#1{{\fontfamily{cmr}\fontseries{bx}\fontshape{it}%
    \selectfont #1}}

  \newmathalphabet{\mathbfit} 
  \addtoversion{normal}{\mathbfit}{cmr}{bx}{it}
  \addtoversion{bold}{\mathbfit}{cmr}{bx}{it}
  \newmathalphabet{\mathbfss} 
  \addtoversion{normal}{\mathbfss}{cmss}{bx}{n}
  \addtoversion{bold}{\mathbfss}{cmss}{bx}{n}
  \ifAMStwofonts
    \ifCUPmtlplainloaded \else
      \UseAMStwoboldmath
      \makeatletter
      \new@mathgroup\upmath@group
      \define@mathgroup\mv@normal\upmath@group{eur}{m}{n}
      \define@mathgroup\mv@bold\upmath@group{eur}{b}{n}
      \edef\UPM{\hexnumber\upmath@group}
      \new@mathgroup\amsa@group
      \define@mathgroup\mv@normal\amsa@group{msa}{m}{n}
      \define@mathgroup\mv@bold\amsa@group{msa}{m}{n}
      \edef\AMSa{\hexnumber\amsa@group}
      \makeatother
      \mathchardef\upi="0\UPM19
      \mathchardef\umu="0\UPM16
      \mathchardef\upartial="0\UPM40
      \mathchardef\leqslant="3\AMSa36
      \mathchardef\geqslant="3\AMSa3E

       \let\le=\leqslant

    \fi
  \fi
\fi 

\ifnfsstwo
  \newcommand{\rmn}[1] {\mathrm{#1}}

  \def\textbfit{\protect\txtbfit}
  
  \long\def\txtbfit#1{{\fontfamily{cmr}\fontseries{bx}\fontshape{it}%
    \selectfont #1}}

  \DeclareMathAlphabet{\mathbfit}{OT1}{cmr}{bx}{it}
  \SetMathAlphabet\mathbfit{bold}{OT1}{cmr}{bx}{it}
  \DeclareMathAlphabet{\mathbfss}{OT1}{cmss}{bx}{n}
  \SetMathAlphabet\mathbfss{bold}{OT1}{cmss}{bx}{n}
  \ifAMStwofonts
    \ifCUPmtlplainloaded \else
      \DeclareSymbolFont{UPM}{U}{eur}{m}{n}
      \SetSymbolFont{UPM}{bold}{U}{eur}{b}{n}
      \DeclareSymbolFont{AMSa}{U}{msa}{m}{n}
      \DeclareMathSymbol{\upi}{0}{UPM}{"19}
      \DeclareMathSymbol{\umu}{0}{UPM}{"16}
      \DeclareMathSymbol{\upartial}{0}{UPM}{"40}
      \DeclareMathSymbol{\leqslant}{3}{AMSa}{"36}
      \DeclareMathSymbol{\geqslant}{3}{AMSa}{"3E}

       \let\le=\leqslant

    \fi
  \fi
\fi 

\ifCUPmtlplainloaded \else
  \ifAMStwofonts \else 
    \def\upi{\pi}
    \def\umu{\mu}
    \def\upartial{\partial}
  \fi
\fi
\def\SV{$\omega_z(t_{\mathrm r})$~}
\def\SVD{$\dot{\omega}_z$~}

\title[Harmonic analysis of CMB data]{Harmonic analysis of cosmic microwave 
background data\\ I: ring reductions and point-source catalogue}
\author[F.~van Leeuwen et al.]
       {F.~van Leeuwen,$^{1,2}$ A.\ D.~Challinor,$^2$ D.\ J.~Mortlock,$^{1,2}$
          M.\ A.\ J.~Ashdown,$^{1,2}$ \newauthor M.\ P.~Hobson,$^2$ 
        A.\ N.~Lasenby,$^2$ G.\ P.~Efstathiou,$^1$ E.\ P.\ S.~Shellard,$^3$ 
         \newauthor D.\ Munshi,$^1$ V.\ Stolyarov$^{1,2}$ \\
        $^1$Institute of Astronomy, Madingley Road, Cambridge CB3~0HA, UK\\
	$^2$Astrophysics Group, Cavendish Laboratory, Madingley Road,
          Cambridge CB3~0HE, UK\\
	$^3$DAMTP, Centre for Mathematical Sciences, Wilberforce Road, 
          Cambridge CB3 0WA, UK}
\date{Accepted for publication in MNRAS: 11 December 2001}

\pagerange{\pageref{firstpage}--\pageref{lastpage}}
\pubyear{2001}

\begin{document}

\maketitle

\label{firstpage}

\begin{abstract}
We present a harmonic model for the data analysis of an all-sky cosmic 
microwave background survey, such as {\it Planck}, where the survey is 
obtained through ring-scans of the sky. In this model, resampling and 
pixelisation of the data are avoided. The spherical transforms of the sky 
at each frequency, in total intensity and polarization, as well as the 
bright-point-source catalogue, are derived directly from the data reduced 
onto the rings. Formal errors and the most significant correlation 
coefficients for the spherical transforms of the frequency maps are preserved. 
A clean and transparent path from the original samplings in the time domain 
to the final scientific products is thus obtained. The data analysis is 
largely based on Fourier analysis of rings; the positional stability of the 
instrument's spin axis during these scans is a requirement for the data model 
and is investigated here for the {\it Planck} satellite. Brighter point sources
are recognised and extracted as part of the ring reductions and, on the
basis of accumulated data, used to build the bright-point-source catalogue.
The analysis of the rings is performed in an iterative loop, involving
a range of geometric and detector response calibrations. The geometric
calibrations are used to reconstruct the paths of the detectors over the sky
during a scan and the phase offsets between scans of different detectors;
the response calibrations eliminate short and long term variations in detector
response. Point-source information may allow reconstruction of the beam
profile. The reconstructed spherical transforms of the sky in each frequency
channel form the input to the subsequent analysis stages. Although the methods 
in this paper were developed with the data processing for the Planck satellite
in mind, there are many aspects which have wider implementation possibilities,
including the construction of real-space pixelised maps. 

\end{abstract}


\section{Introduction}

The traditional method used to analyse the continuum distribution of radiation
on the sky is through pixelisation: the calibrated measurements are projected
onto pixels of a sky map, and further analysis of the data is done by means
of the pixelised responses. These methods, which are most powerfully
explored in the HEALPix software \cite{G2}, have the advantage of
providing a visual link to the reductions. Another already well-developed
example of pixelisation is the IGLOO scheme developed by Crittenden \& 
Turok~(1998) and Crittenden~(2000). A properly designed pixelisation
scheme can, in addition, simplify the further analysis, allowing the use of
fast spherical transform algorithms, and removing contaminated regions of the
sky (G{\'o}rski, 1994; Mortlock, Challinor \& Hobson, 2001). A problem with any
pixelisation of survey data is the loss of information due to the binning of
the data. As a result of coverage variations, different pixels have different
formal errors which complicates the application of fast spherical transforms.
Pixels may be correlated with neighbouring pixels, which further complicates
the analysis. The effective beam profile is unlikely to be circularly
symmetric (as a result of intrinsic asymmetries as well as due to unconvolved
sampling intervals) and requires the observations to be deconvolved. The 
effective two-dimensional beam profile
applicable to a pixel therefore depends on the orientations of the  scans
which have contributed to it, information which is complicated to maintain.
Thus, unless a range of `parallel maps' with supplementary information is kept
(the implementation of which would do away with most of the advantages of the
pixelisation), a pixelisation of the data will inevitably lead to loss of
information. This loss affects our ability to recover the statistical
properties of the reduced data, and as such should be considered a potentially
serious disadvantage.

The products of a survey mission like {\it Planck} 
(see for example http://astro.estec.esa.nl/planck) or {\it MAP} 
(http://map.gsfc.nasa.gov/m$\_$mm.htm) are a point source catalogue and a set of 
frequency maps, from which maps of astrophysical components are derived. 
Methods for the component separation have been developed amongst others by 
Hobson et al.~(1998), Prunet et al.~(2001) and Baccigalupi et al.~(2000). In 
the latter paper an independent component analysis (ICA) algorithm is used to 
separate statistically independent signals from the frequency maps. In
Hobson et al.~(1998) component separation on small rectangular fields is
investigated using a maximum entropy technique in the Fourier domain. The 
latter method has been further developed into a full sky implementation at the
resolution required for the {\it Planck} survey \cite{S2}, with the frequency 
and component maps analysed in the form of their spherical transforms. The 
analysis of the cosmic microwave background (CMB; one of the components 
separated) depends ultimately on the reliability of the separation process. 
A major aim of the data analysis should be to derive the input for the 
component separation, the spherical multipoles of the frequency maps, in the 
most reliable way, preserving as much information as possible on formal errors 
and their correlations.

The analysis of data from an all-sky survey mission in which the survey is
built up from circular scans can be carried out efficiently in three steps, as 
was done for the Hipparcos reductions (Lindegren, 1979; ESA, 1997): the first 
step reduces the raw measurements to ring data; the second step combines the
ring data obtained over the mission to reconstruct the sphere; the third step
analyses the sphere data. The {\it Planck} mission fits this model very
well: data is obtained over 1 hour intervals during which the spin axis of
the satellite remains in a nominally `fixed' position. The collection of
scientifically useful data starts at the end of a pointing manoeuvre and 
finishes with the start of the next manoeuvre, which moves the
spin axis to the next scan position. During this interval, which is referred to
as a time-ordered period (TOP), the satellite performs some 60
revolutions. The area of the sky covered by a detector during a TOP is referred
to as a `ring'. The spinning of the satellite causes every detector $d$ to
describe a small circle on the sky, each with its own specific opening
angle $\alpha_d$ (the opening angle is equivalent to the co-latitude of
the ring as seen from the spin-axis position). A slight wobble of the spin
axis (nutation) widens the rings, but this effect is small relative
to the beam width. Data in a ring are referred to the circle defined by
the mean spin-axis position of the satellite during the TOP and the opening 
angle of the detector concerned. The nutation effect will simulate a 
small-amplitude periodic variation of the opening angle for the detector, 
which will be most noticeable for point sources close to the rings as observed 
by the three highest frequency channels. Over a one-year mission 
almost 9000 rings will be produced for each of the 48 High Frequency 
Instrument (HFI) and 56 Low Frequency Instrument (LFI) detectors.

In the harmonic model the three data analysis steps can be summarised as
follows:
\begin{enumerate}
\item The analysis of the measured samplings per individual detector, 
collected over each TOP. The samples are referred to phases along reference 
circles of which the normal through the centre corresponds with the mean 
spin-axis position over the TOP. After cleaning these data from spikes 
(primarily resulting from very high energy radiation and particles), the
phase-ordered data is collected in phase bins for further treatment. The
short-term response variations are derived, using the accumulated differences
between the mean responses in each phase bin and the individual contributions.
After applying the response corrections, the mean sample values per bin are 
re-calculated and the data is ready for analysis. The first step in the 
analysis is the point-source transit identification. This is initially carried
out on the phase-binned data, but in later iterations the point source 
catalogue constructed from the data is also used. The entire phase-binned 
signal is then fitted with Fourier components for the background signal and 
corrections to the assumed abscissae and intensities for all identified point 
sources. The corrected abscissae and intensities, with their formal errors, 
form the input to the point-source catalogue construction. The transit 
profiles of the brightest point sources are accumulated to provide the input 
for the central-beam-profile calibration.
\item Joint analysis of the Fourier components of the rings from all detectors 
operating at the same frequency, to produce the spherical transforms of the 
frequency maps. For each ring and each detector there is a unique coupling 
matrix, which describes how the spherical multipoles are projected on a small 
circle with a given opening angle $\alpha_d$ and a two-dimensional beam 
profile. The accumulated coupling matrices and ring harmonics for a given 
frequency are the input to a generalised least squares solution for the 
multipoles of that frequency map. In an iterative process, the long-term 
response variations are calibrated; calibration of the outer beam profile may 
also be done at this stage. The processing up to this point is very similar 
for scalar and polarised data \cite{C1}.
\item The analysis of the spherical multipoles of the frequency maps (for 
example, by means of the maximum entropy method) to extract the faint point 
sources, the Sunyaev-Zel'dovich clusters (Sunyaev \& Zeldovich, 1980) and the 
spherical multipoles of the component maps, which are further analysed in the 
form of their power spectra.
\end{enumerate}
Due to the way these steps are interlinked with geometric and response 
calibrations and the construction of the point source catalogue, they will 
require iterations before reduction results can be considered satisfactory.

The present paper focuses on the the first of the three steps mentioned above.
Although this is seen here as the first step in the harmonic model for the
data analysis, many considerations made in the reductions as described here
are equally applicable in a more traditional pixelised approach. The second 
step is described in the accompanying paper \cite{C1}, and covers the 
reductions of the ring data to spherical multipoles of frequency maps in 
detail, a process which is more specific to the harmonic data model. The third 
step, described by Stolyarov et al.~(2002), is the full-sky maximum entropy 
component separation. Additional aspects concerning the analysis of incomplete
sky maps are discussed in Mortlock et al.~(2002).

An important criterion for the implementation of the harmonic model is the 
stability of the spin-axis pointing. This stability is
affected by two main contributors: residual velocities around the two axes
perpendicular to the spin axis, which will create nutation; and external
torques, which can create residual velocities. The amplitudes of the
residual velocities at the start of a TOP are determined by the criteria
for the nutation damping, and should be small enough not to create a
significant effect (in comparison with the beam width) on the accumulated
data. The nutation wobble of the spin axis and the effect of the external
torques can be derived from an analysis of the satellite's attitude in
analytic form and through numerical integration. An analysis of this kind
for the {\it Planck} satellite is presented in Appendix~\ref{App_att}, the
results of which are summarised in Section~\ref{Sec_att}.

In Section~\ref{Sec_calibr} we show how the data analysis in the harmonic 
model is closely linked with geometric and response calibrations.
Details of the data analysis method are presented in Section~\ref{Sec_circ}.
Section~\ref{Sec_point} describes the construction of the point-source 
catalogue and the calibration of the reference phases for the reduced TOPs.
Section~\ref{Sec_geom} presents the geometric calibrations: the focal plane 
geometry, the opening angle correction and the focal plane orientation 
correction. Finally Section~\ref{Sec_iterate} describes the way the data 
reduction and calibration processes are linked in iterative loops.

\section[]{Attitude analysis for the \textbfit{Planck\/} mission}
\label{Sec_att}

The results of a full analytic and numerical analysis of the {\it Planck} 
satellite attitude are presented in Appendix~\ref{App_att}. Here we summarize 
those aspects which are of immediate importance for the data analysis. Though 
this analysis is applied to parameters associated with the {\it Planck} 
satellite, it can easily be modified to apply to different satellite 
configurations, as long as the outer product of the rotational velocity and
angular momentum vectors is high with respect to the external torques acting 
on the satellite.

The {\it Planck} satellite is designed for a survey mission: over a period of 
a year the entire sky will be scanned twice, providing maps of the microwave 
sky at pass bands with central frequencies between 30~GHz and 857~GHz. The 
scanning will take place at pre-determined nominal pointings of the spin axis 
(pointing in a roughly anti-Solar direction). These nominal positions will not 
be exactly reproduced: a pointing noise of a few arcmin is expected. The 
satellite will rotate at a nominal velocity of 1~rpm, but small variations are 
expected here too. The interval of $\sim 1$~hour between two re-positioning 
manoeuvres of the spin axis is referred to as a TOP. During 
a TOP, the satellite's spin axis will describe a relatively stable nutation 
ellipse in the satellite coordinate system (for as long as it remains 
unaffected by nutation damping), the maximum amplitude of which is 
determined by the residual velocities around the $x$- and $y$-axes, and should 
be less than 1.5~arcmin at the end of the nutation damping. The period of the 
nutation will be around 3~to 4~minutes, and will slowly change over the 
mission due to changes in the inertia tensor. The noise on the movement 
of the spin axis of the satellite relative to the nutation ellipse is at the 
arcsecond level. Such variations are not relevant for detectors with a 
5~arcmin or more \textsc{fwhm} response beam. This allows one to incorporate 
a simple set of nutation ellipse parameters in the data analysis if necessary.

\begin{figure}
\centerline{\psfig{file=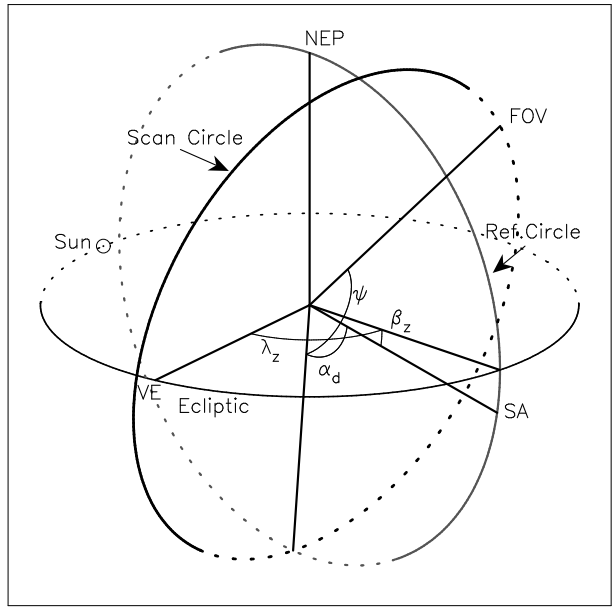,width=8truecm}}
\caption{The orientation angles in the {\it Planck} scan. The spin axis 
position (SA)
is defined by its ecliptic coordinates $(\lambda_z,\beta_z)$, measured from 
the vernal equinox (VE); the spin axis position and the North Ecliptic Pole
(NEP) define a meridian, called the `reference circle'; the position of a
ring is defined by its opening angle $\alpha_d$; the scan phase $\psi$ 
defines the angle between the fiducial reference point (FRP) in the field of 
view (FOV) and the reference circle.}
\label{fig_ang}
\end{figure}
The rotation of the satellite will cause every detector $d$ to describe a 
small circle on the sky, at an angular separation $\alpha_d$ 
(referred to as the opening angle) from the position of the spin axis (see 
Fig.~\ref{fig_ang}). As the actual spin axis will not coincide exactly with 
the nominal spin axis as defined in the hardware, there will be a difference 
between the actual opening angle and its hardware specification value. The 
wobble of the spin axis slightly widens the ring, but for most detectors the
beam width is such that this can be ignored. This may not be the case for the 
highest frequency detectors, where the spin axis wobble may add a little to 
the beam width as observed perpendicular to the scan direction. The way it 
does so will be unique to each ring set due to the variable conditions at the 
end of nutation damping (see Appendix~\ref{App_att}).

We can conclude that, within the model presented, the attitude of the satellite
can be described in a simple manner, while the actual attitude motions
do not cause any problems for the harmonic analysis of the data. The chance of
significant unaccounted forces acting on the satellite, which could 
disturb this situation, is small.

\section[]{The harmonic data model and system calibrations}
\label{Sec_calibr}

The harmonic data model aims to provide an accurate analysis of the data
and its statistical properties. It does so through a sequence of reduction 
steps, resulting in frequency maps in the form of spherical multipoles with 
their covariance matrices and a list of point source positions and intensities
with associated formal errors. These maps can then be used as input for the 
component separation process, which can be performed in spherical multipole
space \cite{S2}. It should be possible to derive the spherical transforms of 
the frequency maps from the Fourier components of the rings, using information 
on pointing, focal plane geometry, focal plane orientation, the opening angle,
the noise correlations of the data and the beam profile, although a number of 
computational problems makes this stage non-trivial \cite{C1}.

The {\it Planck} mission is designed to obtain full-sky surveys at nine
different wavelengths. It explores new ranges of resolution, coverage and 
sensitivity at these wavelengths. For such pioneering observations, 
calibrations are best done by demanding internal consistency of the observed 
sky, which in most cases will not have changed from one observation to the 
next. The exceptional variable and moving sources should all be identifiable. 
Demanding internal consistency defines simple, differential reduction 
procedures, and should result in mission products with the highest possible 
internal precision. Absolute calibrations are carried out only in the final 
stages of the data processing, using a selection of sources that can be 
assigned and measured in a well-controlled way with ground-based or other 
suitable experiments. It is important during the absolute calibrations to 
consider the spectral response of each source and its convolution
with the {\it Planck} passbands, and any background contributions which may be
differently perceived by different instruments. The latter applies in 
particular to calibration sources close to the galactic plane.

The preference for an internal, differential calibration applies to both 
responses and geometric properties. For the responses it involves the 
following steps:
\begin{enumerate}
\item The short-time response variations (time scales from one minute to one 
hour) are calibrated as part of the ring reductions, using the phase-binned 
data (see Section~\ref{Sec_phase}). They show up by displaying, as a function 
of time, the differences between the mean samples per bin and the individual 
contributions to each bin. Removing any systematic effects observed in these
residuals reduces all data contained in one TOP to an arbitrary mean response 
level for a specific detector and ring. All variations at shorter 
time-intervals are reduced to noise contributions, some of which may be 
correlated between neighbouring samples. This is taken into account in the
noise correlation matrix for the model-fit solutions.
\item The long-term response variations (with time scales from one hour to one
year) are calibrated as part of the sphere reductions. The reconstructed 
spherical transforms will, in the first instance, represent an average of all 
mean ring responses. By relating (in Fourier space) the response of each ring
to this mean sky, response corrections per ring and detector are obtained.
This is the equivalent of de-striping \cite{R1}, though in the method
proposed here all data on the ring will contribute. In case the instrument 
high-pass filters the signal, and there is essentially no information left
on the monopole, these adjustments can still be done on the remaining signal.
The response-corrected rings can be used again in a reconstruction of the
spherical transforms, which should result in a decrease in the fitting noise. 
The process can be repeated until no further significant changes
to the ring responses are observed. Given the expected large number of
rings involved (9000 per detector), this process is expected to converge 
quite rapidly. The Hipparcos reductions have demonstrated the validity of
this approach in the iteration between the great circle reduction, the sphere 
reduction and the improvements to the astrometric parameters, which was 
based entirely on internal consistency of the data (see for example
van Leeuwen, 1997). External information may be used to obtain a reasonable
first estimate of the calibration parameters to ensure a rapid convergence
of the process. 
\item The last step in the response calibration adjusts each final 
spherical transform of a frequency map using a set of ground-based (or other) 
calibration sources (see above), thus transforming them from relative to 
absolute scales.
\end{enumerate}
The result of these processes is a set of frequency maps whose
{\it precision} is entirely determined by the internal calibrations, and
{\it accuracy} by the quality and quantity of external calibration 
sources. At any 
stage, should improvements in external sources become available, the external 
calibration can be performed again. In addition, as no external information is 
used in the internal calibrations, the loss of information from the original 
data is minimal. The loss is determined primarily by the limitations in the 
modelling of detector response variations. If instead, for example, properties
of the dipole are used in the earlier calibration steps, the extraction of 
independent information on the dipole from the {\it Planck} data could be 
compromised. The accuracy of the calibration parameters would depend on 
the local amplitude of the dipole signal as projected on a ring, which varies 
strongly with the ecliptic longitude of the spin axis. 

The response or beam profile forms the link between the detector response and 
the detector pointing. The calibration of the central part of the beam profile 
is closely linked with the geometric calibrations and the construction of the 
bright point source catalogue.
The geometric calibrations determine the pointing of each detector at any time
during the observations. The relation between pointing and
time can depend on the following elements:
\begin{enumerate}
\item The ecliptic coordinates of the mean position of the spin axis during a 
TOP: $(\lambda_z, \beta_z)$ (see Fig.~\ref{fig_ang}). This position and the 
position of the North ecliptic pole define a great circle, referred to as the 
`reference circle'. 
\item  For each TOP the nutation ellipse parameters $t_0$, $\omega_{x,0}$, 
$\omega_z$, $f_1$ and $f_2$ as defined in equation~(\ref{eq_om_t0}) in 
Appendix~\ref{app_wobble}. The first two of these determine the $x$-amplitude 
and phase of the nutation ellipse, the other three determine the period and 
the ratio between the $x$ and $y$ amplitudes. The nutation ellipse is 
described in the satellite coordinates of the inertial reference system, as 
defined in Appendix~\ref{App_att}. When observed on the sky, the ellipse
is itself rotating at the spin velocity of the satellite. The nutation period 
changes slowly due to variation in $f_1$ and $f_2$ resulting from depletion
of consumables on-board the spacecraft. 
\item The spin velocity $\omega_z$ and its linear dependence on time during 
a TOP. The latter is again a parameter which changes very slowly over the 
mission.
\item For each TOP, the time of the first transit of the fiducial reference 
point (FRP) in the focal 
plane through the reference circle. A first approximation 
can be obtained from, for example, the star mapper data and the calibration of 
the star mapper\footnote{A star mapper is a device recording the transits of 
bright stars for the purpose of spacecraft attitude control.} position with 
respect to the focal plane geometry. Corrections to the time of first transit 
are obtained during construction of the point-source catalogue.
\item The angle between the FRP and the mean spin axis position (the opening
angle $\alpha_0$, where the index `0' indicates the FRP rather than a specific 
detector `$d$'). This is a correction to the ground-based calibration value, 
that will change slowly with time due to inertia tensor changes.
\item The focal plane geometry, which describes, for an assumed orientation
of the focal plane assembly, the coordinates of all detectors as projected
on the sky, relative to the projection of the FRP in the
focal plane.
\item The focal plane orientation correction, describing the difference
between the actual and the assumed orientation of the focal plane as a 
function of time.
\end{enumerate}
While items (i), (ii) and (iii) are probably best derived from the star 
mapper data, items (iv) to (vii) rely entirely on information contained in the 
science data, in particular on transits of bright point sources. Starting 
values for the focal plane geometry will be obtained from the instrument 
design specifications, but still need to be verified in flight. The 
calibration of these parameters relies on observed abscissae and intensities 
of point sources for different detectors. The geometric calibrations are 
crucial to the reliability of the final mission products, and inevitably 
require a number of iterations through the data reduction and calibration 
processes. As a part of these iterations the bright point-source catalogue 
(BPSC from here on) is produced\footnote{This is not the same as the early 
release point source catalogue to be produced by {\it Planck}, which is 
intended to be an early compilation of approximate coordinates and intensities 
of point sources detected by {\it Planck} to be used in follow-up observations 
by different instruments}. The BPSC 
provides the best possible relative positions and intensities of all brighter 
point sources detected by {\it Planck}, and is finally used to calibrate the 
overall positional reference system to the International Celestial Reference 
System (ICRS; Kovalevsky et al.\ 1997).

With the development of the BPSC, the calibration of 
the central beam profile can gradually be improved. In the first iteration 
through the data it is necessary to assume that all point sources pass through 
the centre of the beam. If, to a first approximation, the beam is represented 
as a two-dimensional Gaussian, then this assumption will do no harm:
the width of the beam profile is in this case not a function of the ordinate 
of the transit. During subsequent iterations the ordinates of the transits can
be calculated using the geometric calibrations and point-source coordinates 
contained in the catalogue. Ordinate information can also be incorporated in 
the beam profile calibrations. The construction of the catalogue ensures 
that in the final iteration a complete sample of point sources (up to a 
maximum ordinate depending on the beam profile) are taken into account in the 
ring analysis. This is essential for the use of such data in the sphere
analysis \cite{C1}. 

The iterations through the data reductions are essential due to the fact
that many of the instrument calibration parameters have to be derived from 
the science data itself. This applies to both the response and the geometric
calibrations. The precision with which these parameters can be determined
is an essential part in the final data-quality verification. Poorly 
determined parameters will inevitably leave their mark on the final data
products. The data analysis needs to identify any such parameters and their
possible effect on the data as part of its preparation for the scientific 
exploration.

\section[]{The ring analysis}
\label{Sec_circ}

The inputs to the ring analysis are the samples collected during a TOP 
for a single detector, supplemented by timing information (an absolute time
for the first sample and a sampling length). It is assumed here that these 
intervals can be recognized from the satellite's housekeeping data. The 
reference position of the spin axis during a TOP is defined as the 
centre of the nutation ellipse actually described by the spin axis, and is 
derived from the star mapper data processing. Each detector $d$ describes a 
small reference circle with opening angle $\alpha_d$ around the 
reference spin axis position.

The use of rings as an intermediate step in the data analysis of a full-sky
survey mission was first proposed by Lindegren (1979) for the Hipparcos 
mission. The idea was independently explored for CMB surveys by 
Delabrouille, G{\'o}rski \& Hivon (1998), who investigated the 
recovery of the CMB power spectrum directly from the ring data.. 

\subsection[]{Signal representation}
\label{Sec_signal}

The signal for the sky can be considered as consisting of three components: a 
continuous background, extended sources (which are not separated from the 
background) and point sources of various intensities. Translating this signal
into the data observed on a given ring, a number of effects have to be taken 
into account:
\begin{itemize}
\item The satellite is rotating at a scan velocity which is not a fixed value
and which may in addition change slightly during a TOP; 
\item The sky signal is convolved with a beam profile and sampled;
\item The sky signal has a (not necessarily constant) background signal added 
to it, originating from the instrument itself; 
\item The sky signal is represented on an arbitrary and possibly slightly 
variable readout scale. 
\end{itemize}
Some of these effects are removed in the ring reductions, others in the 
construction and calibration of the harmonic frequency maps \cite{C1}.

In the analysis of a TOP, two types of contributions are solved for:
\begin{enumerate}
\item The Fourier components, representing the structure in the underlying 
continuum of the microwave sky and extended sources: $C_0$, $C_n$ and 
$S_n$, with $n=1,\dots,{n_{\mathrm max}}$; 
\item The intensities and abscissae for point sources (ordinates are assumed
zero or obtained from the point-source catalogue): $I_k$ and $\psi_k$, with 
$k=1,\dots,s$, $s$ being the number of sources solved for.
\end{enumerate}
In addition, the following properties of the data are resolved:
\begin{enumerate}  
\item The detector response variations, such as changes in the background 
signal, $\Delta b(t-t_{\mathrm r})$, and drifts in the quantum efficiency, 
$\Delta q(t-t_{\mathrm r})$, where $t_{\mathrm r}$ is an arbitrary reference 
time;
\item The covariance matrix of the measurement noise. This will be the noise 
on bin-averaged data rather than individual samples (see 
Section~\ref{Sec_phase}). 
\end{enumerate}
The Fourier components for all TOPs corresponding to detectors 
in the same frequency channel are used to reconstruct the spherical multipoles
of the frequency maps \cite{C1}. This complements the ``ring-torus''
methods developed by Wandelt \& Hansen (2001) for power spectrum estimation 
from the analysis of ring data from a special class of scans. The methods 
required for harmonic map-making also draw heavily on the harmonic-space
convolution algorithms developed by Wandelt \& G{\'o}rski (2001) and 
Challinor et al.~(2000). The ring analysis is iterated with the construction
of the BPSC (see Section~\ref{Sec_point}) by means of
the geometric calibrations mentioned in Section~\ref{Sec_att}. The complete 
inclusion of identified point sources is also ensured through the BPSC.
The success of this iteration determines the final pointing noise 
uncertainties and their contributions to the noise on the frequency maps.
\subsection[]{The time-to-scan-phase relation}

The first requirement for the processing of the TOP data is an accurate 
determination of the scan velocity and its change with time over the interval
covered by the TOP. The scan velocity $\omega_z(t)$ determines the relation 
between time $t$ and a relative scan phase $\psi$ for the individual sampling 
intervals:
\begin{equation}
\psi(t) = \psi(t_{\mathrm r}) + \int^t_{t_{\mathrm r}}\omega_z(t){\mathrm d}t.
\end{equation}
The attitude simulations (Appendix~\ref{App_att}) show that the scan velocity 
can to high accuracy be approximated by the scan velocity at 
reference time $t_{\mathrm r}$, \SV, and a constant scan-velocity drift, \SVD,
as
\begin{equation}
\omega_z(t) = \omega_z(t_{\mathrm r}) + (t-t_{\mathrm r})\dot{\omega}_z,
\end{equation}
giving the following relation for $\psi(t)$:
\begin{equation}
\psi(t) = \psi(t_{\mathrm r}) + (t-t_{\mathrm r})\omega_z(t_{\mathrm r}) + 
\frac{1}{2}(t-t_{\mathrm r})^2\dot{\omega}_z.
\label{equ_psi}
\end{equation}
A further disturbance on the time to scan phase relation comes from the wobble
of the spin axis, but this is very small as is shown in 
Appendix~\ref{app_wobble}. 

The scan velocity and its variation will most probably be derived from the 
star mapper data. However, Section~\ref{Sec_PSPar} shows how information on 
the scan velocity is, in principle, also present in the
science data. There may occasionally be a discontinuity in the relation 
between time and scan velocity due to semi-discrete torques caused by the 
satellite being hit by a micrometeorite. 

Using equation~(\ref{equ_psi}), phases can be assigned to each sample within 
the TOP. Sorting the data explicitly or implicitly (through a reference 
index) on $\psi$ modulo $2\pi$ produces the phase-ordered data (POD) for 
a TOP.  The use of POD has also been explored by Wandelt \& Hansen
(2001) in the context of power spectrum estimation. The further processing of 
the POD involves the following steps, which are described in detail in the 
sections below:
\begin{enumerate}
\item Spike detection is performed on the POD, where spikes will show up more 
clearly due to the $\sim 55$ times higher density of data points 
(Section~\ref{Sec_spike});
\item Phase binning of the data: This achieves a very significant compression 
of the data without significant loss of information (Section~\ref{Sec_phase});
\item Corrections for short-term detector response variations 
(Section~\ref{Sec_ShResp});
\item Point source identification: either from the data stream itself or
from the point-source catalogue, providing abscissae, ordinates 
and intensities (Section~\ref{Sec_PSPar});
\item Signal fitting: solving for the Fourier components representing the 
continuum and corrections to some of the point-source parameters, and the
noise spectrum (Section~\ref{Sec_CirSol});
\end{enumerate}
Processes (i), (ii) and (iii) are applied only once to the data, while 
processes (iv) and (v) are part of the iteration with the BPSC construction.

\subsection[]{Spike detection and removal}
\label{Sec_spike}

In the POD spikes will be much more conspicuous than in the TOD, as data 
points are compared with others at almost identical telescope pointings. 
Filters have to be developed that can reliably detect spikes as outlying 
points. All spikes are to be removed, while their times and intensities should 
be collected to allow for tests of their statistical properties (distribution 
over time and intensity). It may also be necessary to remove the sample(s) 
immediately following a spike in the TOD, and to compare observed response 
variations for detectors with the occurrence of spikes. Confusion between 
spikes and transient objects should be carefully avoided. Once the POD has 
been searched for, and cleaned from, spikes, it is ready for further analysis.
 
\subsection[]{Phase binning of the ring data}
\label{Sec_phase}
 
A typical TOP for an HFI detector will contain around $7\times 10^5$ samplings 
(200~Hz sampling frequency). The Fourier analysis is likely to require an 
$l_{\mathrm max}$ value of around 2500, giving some 5000 unknowns (see 
Section~\ref{Sec_resol}). The process of phase binning compresses the ring 
data without significantly modifying it (as would be the case when resampling),
and as a result does not inflict a significant loss of information. The 
compression of the data, typically by a factor of 40 to 50, significantly 
reduces the processing time and data storage requirements. Relative to the 
original observations, the phase-binned data provides a much improved basis
for the detection of point sources and detector response variations. The phase
binning works for both point sources and the harmonic analysis of the 
background signal and is based on principles developed for, and used 
extensively in, the Hipparcos data analysis of the 1200~Hz modulated main 
detector signal (ESA, 1997; van Leeuwen, 1997).

The basic relation between an observation (one sampling $O_i$ by a single 
detector) and its Fourier representation is given by
\begin{equation}
O_i = C_0 + \sum_{n=1}^{n_{\mathrm max}}\bigl[C_n\cos n\psi_i + 
S_n\sin n\psi_i\bigr] + N_i,
\label{equ_fou}
\end{equation}
where $\psi_i$ is the phase of the observation and $N_i$ represents the 
instrument noise. We will use this representation to describe the signal
with all bright point sources removed. The ring is divided into $m$ 
phase bins of equal length $2\pi/m$. The phase at the centre of each bin is 
given by $\Psi_j=2\pi j/m$. Every observation is associated with a bin $j$, 
which turns equation~(\ref{equ_fou}) into
\begin{eqnarray}
O_i &=& C_0 + \sum_{n=1}^{n_{\mathrm max}}\bigl[C_n\cos n(\Psi_j+{\mathrm d}
\psi_{ij}) \nonumber \\ 
&&+ S_n\sin n(\Psi_j+{\mathrm d}\psi_{ij})\bigr] + N_i,
\label{equ_fou2}  
\end{eqnarray}
where ${\mathrm d}\psi_{ij}=\psi_i-\Psi_j$. This can be further expanded to
\begin{eqnarray}
O_i & =& C_0 + \sum_{n=1}^{n_{\mathrm max}}\biggl[C_n(\cos n\Psi_j\cos n
{\mathrm d}\psi_{ij}\nonumber \\ &&-\sin n\Psi_j\sin n{\mathrm d}\psi_{ij}) +
S_n(\sin n\Psi_j\cos n{\mathrm d}\psi_{ij}\nonumber\\
&& +\cos n\Psi_j\sin n{\mathrm d}\psi_{ij})\biggr] + N_i.
\label{equ_fou3}  
\end{eqnarray}
\begin{figure*}
\centerline{\psfig{file=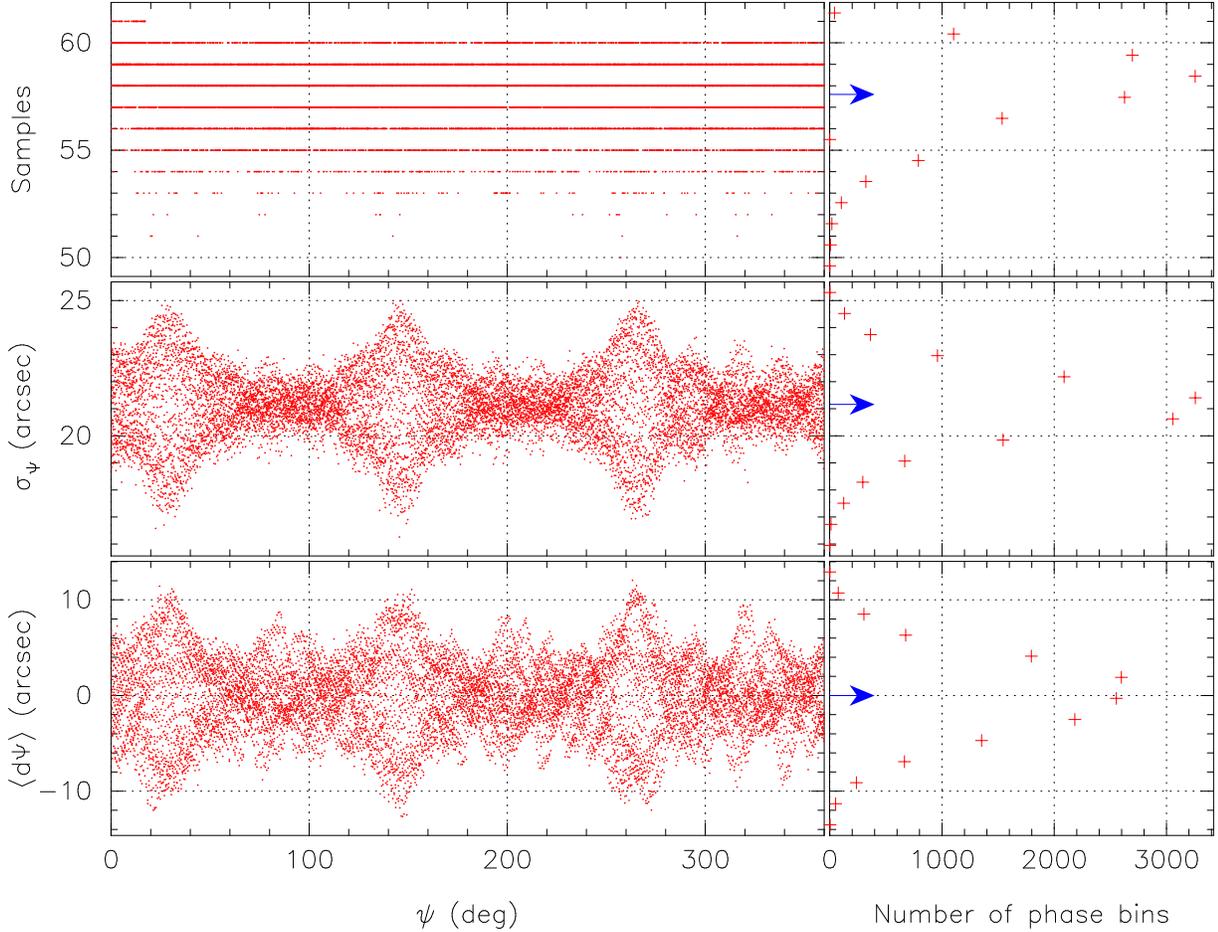,width=16truecm,angle=-90}}
\caption{Results for a binning experiment at the highest {\it Planck} 
resolution: 5~arcmin, using 12500 bins. On the left are shown the actual values
per phase bin as observed, on the right the histogram of the distribution
of observed values, with an arrow indicating the expected mean value.
Top graph: number of samples per bin; middle graph: phase dispersion per bin; 
bottom graph: mean phase offset per bin.}
\label{fig_bins}
\end{figure*}
Phase binning provides weighted means per bin $j$ for the left- and
right-hand sides of equation~(\ref{equ_fou3}). The weights for all 
contributions in a bin can be taken equal. If $n_j$ denotes the number of 
samples in bin $j$, then the following relation is obtained for the mean 
signal in a bin:
\begin{eqnarray}
{\cal O}_j &=& \frac{1}{n_j}\sum_{i\in j} O_i = C_0 + 
\sum_{n=1}^{n_{\mathrm max}}\biggl[\nonumber \\
&&C_n\bigl(
\Sigma_{{\mathrm c},j}\cos n\Psi_j -
\Sigma_{{\mathrm s},j}\sin n\Psi_j\bigr)  \nonumber \\
&&+S_n\bigl(
\Sigma_{{\mathrm s},j}\cos n\Psi_j +
\Sigma_{{\mathrm c},j}\sin n\Psi_j\bigr)\biggr] + {\cal N}_j, \nonumber \\
\label{equ_fou4}
\end{eqnarray}
where $\Sigma_{{\mathrm s},j}$ and $\Sigma_{{\mathrm c},j}$ are defined below.
By choosing the bin size sufficiently small, the sums over the bins can be 
estimated. Experience with the Hipparcos data \cite{E1} showed that a 
coverage of a full cycle of the highest spatial frequency by 6 bins is 
sufficient, and allows for the following approximations to be made:
\begin{eqnarray}
\Sigma_{{\mathrm s},j} &=&\frac{1}{n_j}\sum_{i\in j}\sin n{\mathrm d}\psi_{ij}
\approx  
\frac{n}{n_j}\sum_{i\in j}{\mathrm d}\psi_{ij} = n\langle{\mathrm d}\Psi_j\rangle,\nonumber \\ 
\Sigma_{{\mathrm c},j}&=&\frac{1}{n_j}\sum_{i\in j}\cos n{\mathrm d}\psi_{ij} 
\nonumber \\
& \approx & 1-\frac{n^2}{2n_j}\sum_{i\in j}({\mathrm d}\psi_{ij})^2 = 
1-n^2\sigma^2_{\Psi_j}.
\label{equ_fou5}
\end{eqnarray}
Depending on the amplitude of the nutation ellipse relative to the beam width, 
it may be necessary to fit the data contained in a phase bin as a second-order 
function of the offset from the reference ring position, primarily to avoid 
an intensity bias and noise increase due to point sources. 

The phase binning thus requires the accumulation of four items per bin: the 
mean count ${\cal O}_j$, the total number of samples $n_j$, the average phase 
correction $\langle{\mathrm d}\Psi_j\rangle=\sum{\mathrm d}\psi_{ij}/n_j$ and 
the phase dispersion $\sigma_{\Psi_j}=\sqrt{\sum({\mathrm d}
\psi_{ij})^2/2n_j}$.
Values for $\langle{\mathrm d}\Psi\rangle$ and $\sigma_\Psi$ are shown in 
Fig.~\ref{fig_bins} for a simulation using a sampling frequency of 200~Hz,
${n_{\mathrm max}}=2050$, and 12500 bins. The offset from the nominal 
scan velocity was 1.243~arcsec~${\mathrm s}^{-1}$, and the scan velocity drift 
was 0.009~arcsec~${\mathrm s}^{-2}$. An average of 57 samples per bin 
is expected under these conditions. The expected value for 
$\langle{\mathrm d}\Psi\rangle$ is zero, and $\langle\sigma_\Psi\rangle = 
(\pi/m)/\sqrt{6}$ (where $m$ is the total number of bins), which equals 
$21\farcs 6$, equivalent to 5.8~per cent of the beam dispersion, or 5~per cent
of the bin width in this example. 

The phase-binned data ${\cal O}_j$ and $\langle{\mathrm d}\Psi_j\rangle$, 
$\sigma_{\Psi_j}$ and the correlation matrix of the ${\cal N}_j$ enter our 
model of the response through equations~(\ref{equ_fou4}) and (\ref{equ_fou5}) 
in a generalised least squares solution for the $\{C_n,S_n\}$.

In the evaluation of equation~(\ref{equ_fou5}) we ignored the fourth order 
term, which has an expected value of $(n\pi/m)^4/120$. With $m=6n$ for the 
highest $n$ value used, this amounts to a maximum correction of 
$6\times 10^{-4}$ to $\Sigma_{{\mathrm c},j}$. This figure should be compared 
with the expected amplitudes for the highest frequency harmonics (which will
be severely depressed by the beam convolution), relative to the noise on the 
measurement ${\cal O}_j$. For the lower frequencies the contribution of the 
fourth order term is very much smaller still (e.g.~$6\times 10^{-12}$ for 
$n=25$). In fact, for most of the lower frequencies the approximations of 
$\Sigma_{\mathrm s}\approx 0$ and $\Sigma_{\mathrm c}\approx 1$ can be used. 

Thus, phase binning brings down the number of observations used for the ring 
analysis by a factor $\sim 57$, without any significant loss of information.
The data storage requirements are as a result significantly reduced. All 
trigonometric coefficients in the analysis are fixed in the binned solution, 
and can be pre-calculated. The phase-binned data for each ring can be kept as 
intermediate data products, as the contents do not change in the iterative
processes. 

\subsection[]{Short-term response variations}
\label{Sec_ShResp}

As stated before (Section~\ref{Sec_signal}), the short-term response 
variations can be derived from the systematic differences between the 
individual sample counts and the mean count of the phase bin to which the 
sample has been assigned. Correlated noise between neighbouring samples will
transfer to similar noise between neighbouring bins. Two types of corrections 
can be made: 
\begin{itemize}
\item A background variation correction $\Delta b(t-t_{\rm r})$, which can 
represent effects like changes in the radiation received from the instrument 
and the spacecraft, and which acts like a varying constant offset;
\item A quantum efficiency variation correction $\Delta q(t-t_{\rm r})$, which 
will produce residuals scaled by the mean bin count.
\end{itemize}
For a sample $O_i$ in phase bin $j$ with mean count ${\cal O}_j$, this means
\begin{equation}
\Delta O_{ij} \equiv O_i - {\cal O}_j = \Delta b(t-t_{\rm r}) + 
\Delta q(t-t_{\rm r}){\cal O}_j + N_i.
\end{equation}
The functional representation for these two corrections has to be decided upon 
early in the mission, on the basis of the features shown in the data. A fit 
with a low-order spline function will in most cases take care of any real 
variations. At this stage it is unnecessary, and potentially even damaging, to 
reduce the detector responses to an absolute scale, in particular if those 
calibrations would include a dependence on the phase angle $\psi$, for which 
the reference phase has not been accurately determined yet.

The observed variations should be compared with temperature records for
the spacecraft and payload and the occurrences of spikes, and correlations may 
be used in the removal of any observed variations. As a result of this 
calibration, the data collected during each TOP will be free from short-term 
detector variations. 

\subsection[]{Corrections for the satellite's motion}
\label{sec_aberr}

The {\it Planck} satellite will be positioned in a Lissajous orbit around the 
L2 Lagrangian point of the Sun-Earth system. It will therefore describe an 
almost circular orbit about the Sun with a one year period and a radius of 
$\sim 1.01$~AU. The period for the Lissajous orbit relative to L2 is around 
179 days, and will have an amplitude of around $10^5$~km. Thus, the orbital 
velocity of the satellite is dominated by its motion around the Sun and will 
be approximately 30.3~km~s$^{-1}$. This causes fractional spectral shifts of 
$\Delta\lambda/\lambda\approx 10^{-4}$, which is equivalent to 9 per~cent of 
the dipole 
signal in the CMB radiation. The {\it Planck} mission aims at detecting much 
smaller anisotropies in the CMB, and these effects are therefore a significant 
distortion of the signal. The effect will be opposite in the two  half-year 
surveys, and will be most noticeable near the ecliptic plane.

The data can be corrected for this effect iteratively with the production
of the frequency maps. The frequency maps can be prepared to relatively low
values of $l_{\mathrm max}\approx500$ to produce all-sky spectral index maps. 
The velocity vector of the satellite together with estimated maps of the 
spectral gradient can then provide corrections to the observed intensities 
for each ring:
\begin{equation}
\Delta I \approx \lambda\frac{\partial I}{\partial \lambda}\frac{v}{c}\cos\theta,
\end{equation}
where $\theta$ is the angle between the velocity vector of the observer
(which has magnitude $v$) and the observation direction, $c$ is the speed of 
light in vacuum, $\partial I/\partial \lambda$ is the local spectral gradient,
and $\Delta I$ the local intensity correction. This effect then has to be 
integrated over the spectral response of the beam profile to correct the
actual observed signal.

The positional effect, generally referred to as aberration, is, to first 
order in $v/c$, given by
\begin{equation}
\sin\Delta\theta = (v/c)\sin\theta,
\end{equation}
where  $\Delta\theta$ is the difference between the propagation direction of 
the radiation in a stationary reference frame and the actual moving reference 
frame. For the {\it Planck} observations 
$\Delta\theta$ has a maximum of 20~arcsec (for the part of the scan nearest 
to either of the ecliptic poles) and can in principle be taken into account 
when assigning scan phases to the individual samples. This is likely to be 
relevant for the HFI detectors, for which the maximum correction is comparable 
with the abscissa accuracies for bright point sources, and the angular scale 
of the highest harmonics used in the signal analysis. Ignoring the correction 
will result in information leakage between neighbouring frequencies in the
harmonic analysis, and a significant positional noise contribution for point
sources near to the ecliptic poles.

\subsection[]{Resolution requirements}
\label{Sec_resol}

\begin{figure}
\centerline{\psfig{file=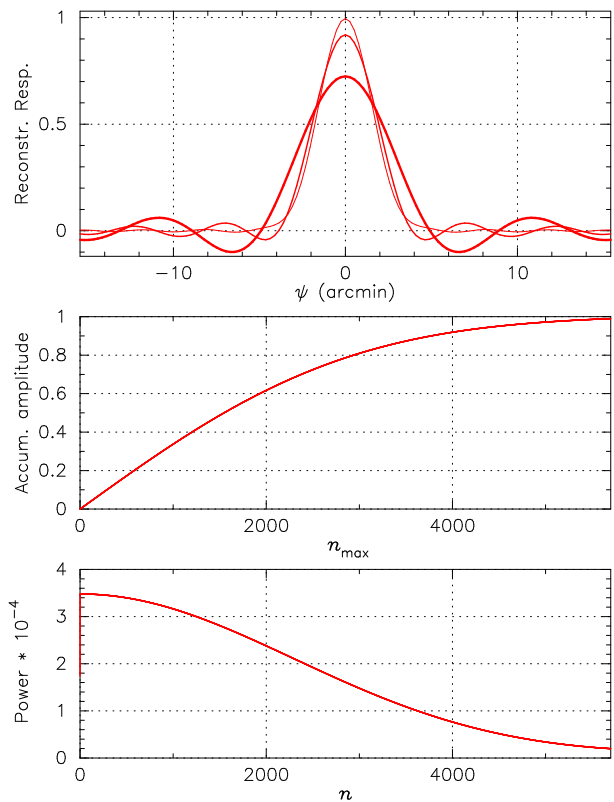,width=8truecm}}
\caption{Bottom graph: The power spectrum of a point source (with sampled peak 
intensity equal to 1) given a Gaussian beam with \textsc{fwhm} of 5~arcmin 
($\sigma_b=2.12$~arcmin) and a sample width of 1.8~arcmin. Middle graph: the 
restored peak intensity of a point source if fitted with harmonic components 
up to the indicated $n_{\mathrm max}$ value. Top graph: the restored point 
source for $n_{\mathrm max}=2500$ (thick), 4000 (intermediate) and 6000 (thin) 
(also distinguished by increasing peak height and decreasing side-lobe 
amplitudes).}
\label{fig_point}
\end{figure}

An important aspect of the realization of the harmonic data model is the value
of ${n_{\mathrm max}}$ that should be applied to the data analysis. This value
is largely determined by the beam width, the scan density, and the way in 
which point sources are dealt with in the analysis. Good estimates for
${n_{\mathrm max}}$ can be obtained by assuming a circular-symmetric Gaussian
beam profile. The power spectrum of a point source convolved with a Gaussian
beam is shown in Fig.~\ref{fig_point}. It is clear that in order to represent 
the point sources adequately as components in the harmonic analysis relatively 
high values of $n_{\mathrm max}$ are required. An alternative approach is to
treat identifiable point sources as separate components in the ring analysis. 
This should be feasible if no features are expected in the 
background signal, which are sharp compared to the beam profile.

A circular symmetric Gaussian beam with a dispersion $\sigma_b$ (in radians) 
will reproduce the higher harmonics in the background signal with decreased 
amplitudes. If the beam is represented by 
\begin{equation}
R(\psi) = \exp[-\psi^2/2\sigma_b^2],
\end{equation}
then the decrease in the amplitude is given by
\begin{equation}
{a_{lm}}' = a_{lm}\exp[-(l\sigma_b)^2/2],
\end{equation}
where ${a_{l}}$ is the actual and ${a_{lm}}'$ the observed amplitude 
(see also Challinor et al., 2002).
Provided that point sources are treated 
as separate objects, the suppression of higher harmonics determines the value 
for $n_{\mathrm max}$ in the ring solution. For $\alpha_d\approx 85\degr$, 
simulations (to be detailed in a future paper) 
suggest a value of $n_{\mathrm max}\approx 4400~{\mathrm arcmin}
/\sigma_b$, where $\sigma_b$ is expressed in arcmin. Thus, for 
$\sigma_b=2.12$~arcmin (\textsc{fwhm} of 5~arcmin) we find 
$n_{\mathrm max}\approx 2100$. At such high $n_{\mathrm max}$ values the 
matrices involved in the transformations are very large, and the covariance 
matrix of the spherical multipole solution will contain around 
$10^{13}$ elements. Fortunately, simulations, which will be detailed 
in a future paper, have shown that for many plausible scanning strategies, the 
covariance matrix of this solution will be very sparse (some analytic 
approximations to the covariance matrices for simple scan strategies are also
derived in Challinor et al., 2001).

\subsection[]{The point-source parameters}
\label{Sec_PSPar}

The main tasks of the point source processing are the following:
\begin{itemize}
\item To identify from either the data stream or the BPSC the point sources 
present in the ring data; the source of 
information will depend on the iteration stage of the data reduction process; 
\item To supplement this list with solar system objects that may have been
observed;
\item To produce for each point source preliminary estimates of the intensity
$I_s$, the abscissa $\psi_\tau$, and if obtained from a catalogue or
ephemerides (solar system objects), the ordinate $\upsilon_\tau$; 
\item To obtain as part of the reduction of the ring data the intensities 
$I_s$ of all identified sources, and abscissae $\psi_\tau$ for sources with a 
sufficiently high signal-to-noise ratio;
\item To collect the abscissa data for (re-)building the BPSC;
\item To collect the profiles and fitting parameters for the brightest 
sources as a contribution to the beam profile calibration.
\end{itemize}
This process is clearly iterative, improving at each step the quality of the
BPSC, the beam profile and the reliability of the segregation of the point 
sources from the data.

\subsubsection[]{Identification of point sources}
\label{Sec_IdentPS}

The mechanism of point source identification will depend on the iteration
phase of the data reductions: in the first instance point sources will mainly
be identified from the data stream itself, while in subsequent iterations the
identification will rely more on the BPSC. It is expected that, at least for 
the high-frequency detectors, the power will be dominated at high $l$ values
by point sources, which should allow for the development of a reliable filter 
for the detection of brighter point sources. Filters for the recognition of 
point sources in a one-dimensional data stream were developed and successfully 
applied for the Hipparcos and Tycho data reductions \cite{L1}. Simplified 
versions of two-dimensional filters under development for recognition of point 
sources on maps (see for example Cayon et al., 2000, Sanz et al., 2001) could 
also be considered.
The detection of point sources from the data is significantly enhanced by the 
phase binning of the data, increasing the signal-to-noise ratio by almost a 
factor 7 for the parameters used in Section~\ref{Sec_phase}. 

On the first pass through the data all point sources are assumed to transit 
through the centre of the beam. This is not problematic if the beam is 
approximately circularly symmetric and Gaussian. The deviation of the actual
beam profile from this assumption will cause a slight error in the first 
reconstruction of the beam. This error can be reduced once 
ordinate information on point source transits becomes available too.

All detected point sources will be used to build the first and subsequent
versions of the BPSC: the evolving catalogue used in iterations to predict and 
consistently identify point sources. In these iterations information on 
the ordinate of the source at the time of the transit should also be 
incorporated. The consistent inclusion of point sources in the ring analysis 
is an essential requirement for the further analysis of the underlying 
continuum. The BPSC also plays a crucial role in the geometric calibrations 
of the instrument.

\subsubsection[]{The measured and binned point-source signal}
\label{Sec_MeasPSSgn}

\begin{figure}
\centerline{\psfig{file=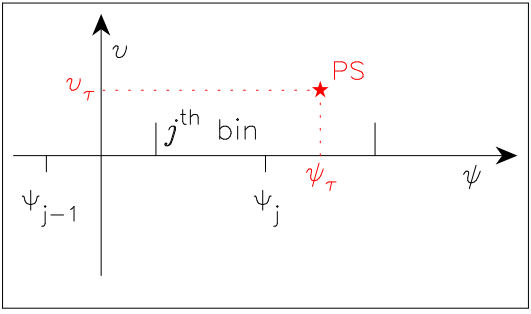,width=7truecm}}
\caption{The point-source position (PS) relative to the phase binning and the 
mean scan circle.}
\label{fig_PS}
\end{figure}
The sampled responses $O_i$ for a point source of intensity $I_s$, passing 
through the beam of a detector, is a function of the scan phase $\psi_i$ (at
the midpoint of the sampling interval) for sample $i$ and the abscissa 
$\psi_\tau$ and ordinate $\upsilon_\tau$ of the point source 
(see Fig.~\ref{fig_PS}):
\begin{equation}
O_i=I_s\int^{\psi_i+\Delta\psi}_{\psi_i-\Delta\psi} 
  R(\psi_\tau-\psi,\upsilon_\tau){\mathrm d}\psi,
\end{equation}
where the integral represents the sampling interval, and 
$R(\psi_\tau-\psi,\upsilon_\tau)$ is the normalized beam profile for a detector
as a function of the offset from the centre of the beam ($\psi_\tau-\psi$ 
along the scan direction and $\upsilon_\tau$ perpendicular to the scan 
direction).
\begin{figure}
\centerline{\psfig{file=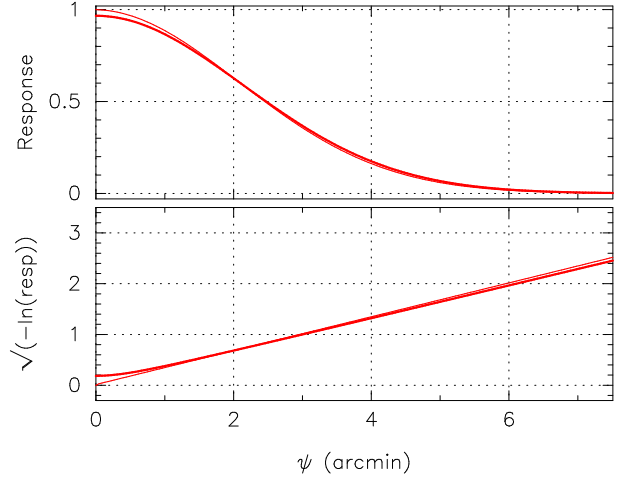,width=8truecm}}
\caption{The beam response $R(\psi,0)$ (thin line) and the convolved beam 
response $S(\psi,0)$ (thick line) profiles for a Gaussian beam with 
\textsc{fwhm}$=5$~arcmin. The increase in beam width as a result of the 
samplingis of the order of 20~arcsec.
Top graph: in linear response scale; bottom graph: in logarithmic response 
scale.}
\label{fig_beam}
\end{figure}
The effective beam profile $S(\psi,\upsilon)$ is defined as the actual beam 
profile $R(\psi,\upsilon)$ convolved with the sampling interval, as 
illustrated in Fig.~\ref{fig_beam} [see also equation~(\ref{equ_fou5})]
\begin{equation}
S(\psi,\upsilon) = \int^{\psi+\Delta\psi}_{\psi-\Delta\psi} R(\psi',\upsilon)
{\mathrm d}\psi',
\end{equation}
where the relevant coefficients are defined in Section~\ref{Sec_phase}.
When phase binning is applied to the point source contributions, the response 
in phase bin $j$ is given by
\begin{eqnarray}  
{\cal O}_j &=& I_s\bigl[S(\psi_\tau-\Psi_j,\upsilon_\tau) -
\langle{\mathrm d}\Psi_j\rangle S^\prime(\psi_\tau-\Psi_j,\upsilon_\tau) \nonumber \\ 
&& + \sigma_{\Psi_j}^2 S^{\prime\prime}(\psi_\tau-\Psi_j,\upsilon_\tau)\bigr] + 
{\cal B}_j + {\cal N}_j,
\label{eq:ps_bin}
\end{eqnarray}
where ${\cal B}_j$ represents the background signal.
In the same way as was found for the harmonic multipoles, the main 
contribution  comes from the second derivative, which produces an effective 
broadening of the beam. However, if we use 12~500 bins for a beam with 
\textsc{fwhm}=5~arcmin ($\sigma_b=2.12~{\mathrm arcmin}$), the additional
dispersion is $\sigma_{\Psi_j}\approx 21\farcs 6$ and the effective
\textsc{fwhm} of the beam is increased as a result of the 
phase binning by no more than one per~cent.

\subsubsection[]{Fitting parameters for the point-source signal}

Two parameters require determination in the signal fit: the transit phase 
$\psi_\tau$ and intensity $I_s$ of the point source. For both, preliminary
estimates are required which are then adjusted in the solution.

Equation~\ref{eq:ps_bin} can be linearized in the intensity correction
$\delta I_s$ and the transit correction $\delta\psi_\tau$ so that
\begin{eqnarray}
{\cal O}_j &=& \delta I_s S(\psi_\tau-\Psi_j,\upsilon_\tau)
 + \delta\psi_\tau I_s S^\prime(\psi_\tau-\Psi_j,\upsilon_\tau) \nonumber \\
&& + I_sS(\psi_\tau-\Psi_j,\upsilon_\tau) + {\cal B}_j + {\cal N}_j.
\label{equ_point}
\end{eqnarray}
This equation enters in the generalised least squares solution for the ring 
data (see Section~\ref{Sec_CirSol}). The intensity correction term in 
Eq.~\ref{equ_point} is scaled by the response function. As a result of this,
the noise on the lower intensities (the wings of the response function) will
tend to affect the determination of $\delta I_s$ more than the better 
determined higher intensities (core of the response function). This is in
particular the case when the noise on the signal depends on its intensity.
Therefore, only a few of the central phase bins should be used for solving 
for the transit parameters of a given source. It may be necessary to iterate 
the ring solution to properly separate the point source and continuum 
contributions.

Assuming we use the central 5 phase bins for determining the fitting parameters
for each point source, estimates can be obtained for the expected precisions.
For the intensity error a value of 0.8 times the noise level on the binned 
samples is expected, equivalent to 0.12 times the noise level on individual
samples. For the abscissa error we estimate a value of 1.7 arcmin divided by 
the signal-to-noise ratio of the point source in the phase-binned data. 

\subsubsection[]{Transient sources}

Any object moving by a significant fraction of the beam width over an interval
of one hour will have to be treated as a transient source. For the highest
frequency channels of {\it Planck} this translates into a movement of 
approximately 2~arcmin per hour and above. The fastest objects in the Earth's 
neighbourhood will be traveling at $\sim 10^5$~km~hr$^{-1}$. The limit of 
2 arcmin then translates into a horizon of around 1.2 AU, which will include 
a small fraction of asteroids and the occasional comet.

For sources that are not variable on a time-scale of less than one 
hour (which may not be true due to rotation of the objects), systematic 
differences between the actual measurements (before phase binning) and the 
reference profile can be expressed as corrections to the effective scan 
velocity \SV:
\begin{eqnarray}
\Delta O_i &=& O_i - B_i - I_sS(\psi_\tau-\psi_i,\upsilon) \nonumber \\ 
&\approx& I_s\frac{\partial S}{\partial\psi}
\Delta(\psi_i-\psi_\tau) + N_i \nonumber \\
&=& (t-t_{\mathrm r})I_s\frac{\partial S}{\partial\psi}{\mathrm d}\omega_z
+ N_i,
\label{eq_phas2}
\end{eqnarray}
which shows that most of the information on \SV is contained in those sections
of the signal which have the steepest gradient as a function of $\psi$. 
The same kind of information can in principle also be used to determine the
scan velocity from the science data using bright-point-source measurements,
although it would be by far preferred to derive this information from the 
star mapper data. 

\subsubsection[]{Beam profile calibration}

The beam profile calibration uses the accumulated transit data of bright point 
sources (which may include transits of solar system objects). During the first 
processing of the data no reliable ordinate 
information is available (catalogue positions are still poor or non-existent, 
and opening angles still need to be calibrated), and every source is 
assumed to go through the centre of the beam. During later stages of the data 
reductions, when the point-source catalogue and the geometric calibrations are
improving, the ordinate information can be incorporated too. 

The principle of the beam profile calibration follows techniques used in the
Hipparcos data reductions for the single slit response functions of the 
star mapper slits (ESA, 1997; van Leeuwen, 1997). The response function is 
sampled at a frequency 4 to 8 times higher than the sampling frequency of the 
data. Using the abscissae and intensities of identified point sources, the 
observed residual samplings ${\cal O}_j-{\cal B}_j$ (original signal 
${\cal O}_j$ minus the background ${\cal B}_j$ as estimated from the current 
harmonic fit to the continuum) in phase bins close to the point source 
transit phases $\{\psi_\tau\}$ are binned according to their separation from 
the transit phase $\{\psi_\tau\}$. Also accumulated in bins are the derived 
transit peak intensities $I_s$. Thus, we find for the beam-profile bin $k$ 
the mean normalised response
\begin{equation}
{\cal S}(k) = \frac{\sum \bigl[{\cal O}_j(k)-{\cal B}_j(k)\bigr]}{\sum I_s(k)},
\label{equ_beam}
\end{equation}
where the sums are over the point sources and phase bins contributing to 
the beam-profile bin $k$.
Noise correlations between neighbouring bins in the phase-sampled data will
also affect the accumulation of equation~(\ref{equ_beam}), in that 
pairs of bins in the accumulation can contain partially correlated noise, 
which needs to be taken into account when fitting a response curve. 
The values ${\cal S}(k)$ can be fitted with a cubic spline to provide a 
smooth beam profile with a continuous derivative, which then provides an
estimate for the sampled beam profile $S(\psi,0)$.
When, after the construction of the first point-source catalogue, ordinates
become available for the point-source transits, the beam profile can be 
resolved in two dimensions. The reconstructed beam profile obtained this way
is the sampled, effective profile, and not the actual profile, which would
apply to a stationary detector.

Transits of planets such as Jupiter may also be useful for the beam profile 
calibration, though could be problematic: the detector response to very 
high intensities will not be linear, and the spectral gradient for the
planets is likely to be quite different from that of the average microwave
point source. The beam profiles will vary with frequency (see for example 
Challinor et al., 2001), making it still more complicated to incorporate
the profiles measured from the planets.

\subsection[]{The ring solution}
\label{Sec_CirSol}

After binning the data, calibrating and removing short-term detector responses,
and identifying the point sources, the TOP is ready for reduction. This part 
of the reductions consists of a generalised least squares solution
for the ring harmonics in the underlying continuum
(equations~(\ref{equ_fou3})--(\ref{equ_fou5})) and simultaneous solutions for 
the abscissa and intensity corrections for all identified point sources
(equation~(\ref{equ_point})). The observation vector $\bmath{z}$ has as
its components the mean response in each phase bin. The observations are
related to the vector $\bmath{x}$ containing the amplitudes of the circle
harmonics, $\{C_n, S_n\}$, and the corrections to the point source
parameters, $\delta I_s$ and $\delta \psi_\tau$, and the noise
vector $\bnu$, whose components are the noise in each phase bin,
${\mathcal{N}}_j$, via the linear equation
\begin{equation}
{\bmath z} = {\mathbfss A}{\bmath x} + {\bnu}.
\label{eq_obs}
\end{equation}
The matrix $\mathbfss{A}$ depends on the locations of the phase bins,
$\Psi_j$, the phase corrections and dispersions, $\langle {\mathrm{d}}
\Psi_j \rangle$
and $\sigma_{\Psi_j}$, the current estimate of the point source intensities
and positions, $I_s$ and $\psi_\tau$ (and ordinate information as this becomes
available), and the current estimate of the beam profile, $S(\psi,v)$.
Equation~(\ref{eq_obs}) is the matrix form of equations~(\ref{equ_fou4})
and (\ref{equ_point}). The minimum-variance estimate of $\bmath{x}$ is
\begin{equation}
\hat{{\bmath{x}}} = ({\mathbfss{A}}^T {\mathbfss{N}}^{-1} {\mathbfss{A}})^{-1}
{\mathbfss{A}}^T {\mathbfss{N}}^{-1} {\bmath{z}},
\end{equation}
with errors 
\begin{equation}
\langle (\hat{{\bmath{x}}} - {\bmath{x}})(\hat{{\bmath{x}}} - {\bmath{x}})^T
\rangle = ({\mathbfss{A}}^T {\mathbfss{N}}^{-1} {\mathbfss{A}})^{-1},
\end{equation}
where ${\mathbfss{N}} \equiv \langle \bnu \bnu^T \rangle$ is the (phase-binned)
noise covariance matrix.

Assuming the instrument noise is a stationary, random process with
correlation function $C(t)$ in the time domain, we can obtain the noise
contribution to the mean response in a given phase bin by integrating
over the time periods corresponding to those observations falling in that
bin. The covariance matrix $[{\mathbfss{N}}]_{jj'}\equiv
\langle {\mathcal{N}}_j {\mathcal{N}}_{j'}\rangle$ then takes the form of a
convolution:
\begin{equation}
[{\mathbfss{N}}]_{jj'} = \left( \frac{m}{N_{\rmn{s}}} \right)^2
\int_{x_-}^{x_+}
\Lambda(x) C\{T_{\rmn{s}}[x + (j-j')/m]\}\, {\rmn{d}} x,
\end{equation}
where $N_{\rmn{s}}$ is the number of times the ring is scanned in one TOP,
$T_{\rmn{s}}$ is the average spin period in that TOP, $m$ is the number
of phase bins, and the integration limits $x_\pm \equiv -(N_{\rmn{s}}-1)\pm
1/m$. The function $\Lambda(x)$ is given by
\begin{eqnarray}
\Lambda(x) &=& \sum_{n=-(N_{\rmn{s}}-1)}^{N_{\rmn{s}}-1}\Bigl[
\Theta(1/m - |x-n|)\nonumber \\
&& \qquad\times(1/m - |x-n|)(N_s-|n|)\Bigr], 
\end{eqnarray}
where $\Theta(x)$ is the Heaviside unit step function, and arises from the
effects of phase binning and repeatedly scanning the ring. For white
noise the ${\mathcal{N}}_j$ are uncorrelated, but in the presence
of a significant low frequency component correlations will arise. Typically,
instruments are designed with the goal of restricting coloured noise to
frequencies below the spin frequency, in which case the correlation length
of the noise exceeds the spin period. (For \emph{Planck} HFI, the
nominal knee frequency at which the character of the noise changes from
$1/f$ to white is less than $0.06\, \rmn{rad}~\rmn{s}^{-1}$, while the
spin frequency is $0.10\,\rmn{rad}~\rmn{s}^{-1}$.) As the correlation
length becomes large compared to the spin period, the noise covariance matrix
approaches $[{\mathbfss{N}}]_{jj'} = \chi^2_{\rmn{c}} + \chi^2_{\rmn{u}}
\delta_{jj'}$, corresponding to a fully correlated offset in every bin with
r.m.s.\ $\chi_{\rmn{c}}$, and uncorrelated noise with r.m.s.\ 
$\chi_{\rmn{u}}$. In practice, the noise power spectrum will have
to be estimated from the data directly rather than relying on simple
parameterised forms like that give above.
If the correlation length exceeds $N_{\rmn{s}} T_{\rmn{s}}$
the offsets on different rings will generally also be correlated, so an
optimal analysis would require reducing several rings simultaneously.

\begin{figure}
\centerline{\psfig{file=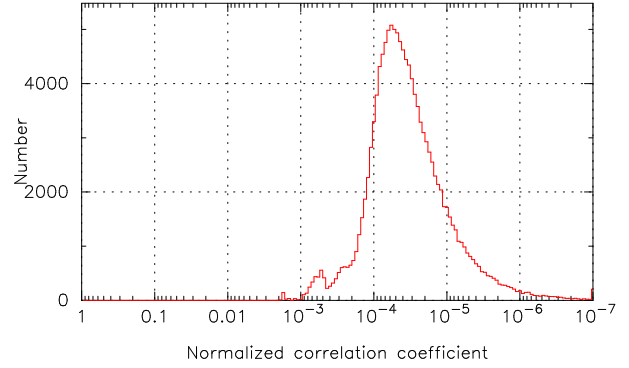,width=8truecm}}
\caption{A histogram of the normalised, absolute values of the
off-diagonal elements in the square root of the covariance matrix for a 
typical bin-sampled ring solution, using $l_{\mathrm max}=512$. The matrix 
can effectively be regarded as diagonal.}
\label{fig_hist}
\end{figure}
Under the assumption of Gaussianity, the relevant components of the
covariance matrix $\langle (\hat{{\bmath{x}}} - {\bmath{x}})(\hat{{\bmath{x}}}
- {\bmath{x}})^T \rangle$ determine the errors on the ring harmonics
(marginalised over the point source corrections). We have conducted several
simulations to investigate the effects of variations in the
spin velocity, and the presence of point sources, on the covariance matrix for
the errors on the ring harmonics. To isolate the effects of spin velocity and
point sources we have only
considered white instrument noise. For our simulations we adopted a
maximum value of $n$ equal to $512$. The covariance matrix for the
errors on the ring harmonics is found to be very 
close to diagonal if only relatively faint point sources are present. In the 
presence of a very bright point source (like Jupiter), which will create a 
very low-weight gap in the ring data, there is a minor effect on the 
diagonal structure of the covariance matrix, the effects being largest in
low frequency detectors. The normalised off-diagonal elements in the 
(Cholesky) square root of a typical covariance matrix are accumulated in a 
histogram in Fig.~\ref{fig_hist}. The results show that the correlations 
between the different ring harmonics are at a level of 0.2~per cent or less.
These levels decrease still further with increased resolution. The inclusion 
of a bright point source would affect the distribution, but correlations
would still be below the 1~per cent level. If we further include stationary
instrument noise, we can expect the errors on the ring harmonics to still
be uncorrelated between Fourier modes for large $N_{\rmn{s}}$.
If coloured noise is confined to frequencies well below the spin frequency,
correlations between rings will only affect the $n=0$ modes~\cite{C1}.

\section[]{The bright-point-source catalogue}
\label{Sec_point}

The bright-point-source catalogue (BPSC) is both a major product of the 
{\it Planck} mission, and an important calibration tool. The catalogue is 
constructed iteratively from the abscissae and intensities of the point 
sources obtained in the current ring reductions. Corrections to the BPSC
will ultimately provide positions, intensities (at different frequencies), 
and variability information for all detected sources. As was explained in
Section~\ref{Sec_calibr}, the spin axis position (including nutation movements)
and spin rate are most likely derived from the star mapper data, all other
geometric calibrations depend at least to some extent on the science data.

\subsection[]{The positional sphere reconstruction}

The construction of the BPSC follows a simplified version of the Hipparcos 
sphere reconstruction process \cite{E2}. This process requires a set of a 
priori positions for the point sources, to which corrections are determined. 
These positions are also required in order to identify consistently 
point-source transits on different rings with objects on the sky. Initial 
positions at a precision of twice the beam size (10~arcmin for {\it Planck}) 
will be sufficient for this purpose. Using the ground-based specifications of 
the focal plane geometry and inertia tensor, first estimates are obtained for 
the opening angle and focal plane rotation corrections, and the focal plane 
geometry. These estimates can be further refined through the use of planetary
transits. From this information we obtain preliminary details of the alignment 
of the different detectors during a scan: phase offsets and effective opening 
angles. The transits recorded in scans for the different detectors can now 
be transferred to a common sky, where each transit is represented by an
error ellipse, strongly elongated in a direction perpendicular to the 
scan direction. A simple algorithm is then required to cross identify the 
different transits, and to produce from the accumulated information the
necessary initial positions.

Once a catalogue of initial positions has been obtained, and transits have 
been 
identified with sources in the catalogue, corrections to the assumed positions
and to the reference phases of the rings can be determined. Given the 
ecliptic coordinates $(\lambda_z,\beta_z)$ of the spin axis, and a position 
for a point source $i$ at $(\lambda_i,\beta_i)$, the expected abscissa and
ordinate of the point source with respect to the ring can be derived.
The coordinates $(\psi_i,\zeta_i)$ in the system with the spin axis at the pole
(where $\psi_i$ is measured from the reference circle as defined in 
Fig.~\ref{fig_ang}, and $\zeta_i$ is the latitude of the source relative to 
the great circle associated with the spin-axis position) are obtained from
\begin{equation}
\left[\matrix{\cos\psi_i\cos\zeta_i\cr\sin\psi_i\cos\zeta_i\cr\sin\zeta_i}\right]
 = {\mathbfss R}_2\bigl(\beta_z-\frac{\pi}{2}\bigr){\mathbfss R}_3(-\lambda_z)
\left[\matrix{\cos\lambda_i\cos\beta_i\cr\sin\lambda_i\cos\beta_i\cr
\sin\beta_i}\right],
\label{eq_coor_tr}
\end{equation}  
where ${\mathbfss R}_i(\phi)$ is the matrix representing a right-handed 
rotation around 
axis $i$ through angle $\phi$. Equation~(\ref{eq_coor_tr}) 
can be used to derive the relation between corrections to the source position
$(\lambda_i,\beta_i)$ and the resulting changes to the scan phase $\psi$ 
(obtained from the transit abscissae) and ordinate $\upsilon$ (reflected in 
the transit intensities).
We define $\zeta=\frac{\pi}{2}-\alpha+\upsilon$, where $\upsilon$ is the 
ordinate relative to the actual ring, and $\alpha$ the opening angle. 
The relations between the offset values for $(\Delta\psi_i,\Delta\upsilon_i)$ 
and $(\Delta\lambda_i,\Delta\beta_i)$ are then found to be
\begin{eqnarray}
\Delta\psi_i\cos^2\zeta_i &=& f_1\Delta\lambda_i\cos\beta_i+f_2\Delta\beta_i, 
\nonumber \\ 
\Delta\upsilon_i\cos\zeta_i &=& -f_2\Delta\lambda_i\cos\beta_i+
f_1\Delta\beta_i, 
\end{eqnarray}
where $f_1$ and $f_2$ are defined as
\begin{eqnarray}
f_1 &=& \sin\beta_z\cos\beta_i-\cos\beta_z\sin\beta_i\cos(\lambda_i-\lambda_z),
\nonumber \\
f_2 &=& \cos\beta_z\sin(\lambda_i-\lambda_z).
\end{eqnarray}
For a scanning strategy where the spin axis remains in or close to the 
ecliptic plane (and $\lambda_i-\lambda_z\approx\pm\pi/2$), $f_1\approx 0$ and 
$|f_2|\approx 1$ for most sources, the 
exception being sources situated close to either of the ecliptic poles. The 
result is that very little information on the ecliptic longitudes can be 
extracted from the measured abscissae, except for images at high ecliptic 
latitudes. Some information on the longitudes can, however, be extracted using 
the measured intensities, which depend on $\upsilon$ through the beam profile.
This method will inevitably fail when the source is variable.

The positional sphere solution consists of a least squares fit of the 
differences between the observed and predicted abscissae, $\Delta\psi_{i,j}$ 
(for point source $i$ on ring $j$) to corrections 
$(\Delta\lambda_i,\Delta\beta_i)$ for each point source. In 
addition, there is a correction $\Delta\psi_j$ to the assumed reference phase
for each ring. In the solution a boundary condition needs to be included which
states that the average correction $\sum\Delta\psi_j=0$, else a singularity 
may occur. When using different detectors, the phase offsets between the 
detectors have to remain effectively fixed: the only changes can come from
rotation variations of the field of view, and focal length variations of the 
telescope, and both effects are unlikely to be significant.  

To estimate the number of observations and parameters, we assume 10~000
point sources on the sky bright enough to be detected on single rings 
(Barreiro, private communication), and 10~000 rings per detector over the 
{\it Planck} mission. With the beam's \textsc{fwhm}$=5$~arcmin, the average 
number of point sources detected per ring is 6 passing within $\pm\sigma_b$, 
and 6 more passing within $\pm2\sigma_b$ from the mean position of the 
ring. The total number of transits recorded is then $\sim10^5$ 
per detector, giving $\sim 10$ observations per source per detector. The total 
number of parameters to be estimated is $\sim3\times10^4$, and this does not 
increase when information from more detectors is used (except when some point 
sources are not visible for all detectors used). Increasing the number of 
detectors increases the number of observations and will increase the rigidity 
of the solution. 

The relative positions of point sources obtained this way are rigid, but 
absolute positions are only determined up to an overall rotation, and so 
require linking to the International Celestial Reference 
Frame \cite{K1}. This can be accomplished through cross identification with 
radio and optical counterparts, thus determining a global rotation to be 
applied to the entire catalogue. This may not be important for the statistical
properties of the CMB component in the background, but is relevant for 
relating sources and the dust map to other observations.

\subsection[]{Geometrical calibrations}
\label{Sec_geom}

In Section~\ref{Sec_calibr} we mentioned four geometric calibrations which 
rely on the point-source data collected by the satellite: 
\begin{enumerate}
\item The reference phase of the FRP, to be obtained for each circle;
\item The opening angle for the FRP, 
to be obtained as a slowly changing function of time;
\item The focal plane orientation, to be obtained as a slowly changing 
function of time.
\item The focal plane geometry, to be obtained as a fixed set of parameters.
\end{enumerate}

The first three points concern the first order geometric orientation of the 
field of view: its shifts perpendicular to, and along, the scan direction, and
its rotation. These result in systematic shifts in the abscissae and blurred 
intensity distributions for point sources, both as functions of the field of 
view position of the detector. The systematic shifts can be solved for as 
instrument parameters in the positional sphere reconstruction. This also
applies to the calibration of the along-scan position of each detector in the 
focal plane geometry.
 
The opening angle for each detector can only be derived from the brightness 
distributions of the point sources observed with it. A maximum likelihood 
solution, optimizing the intensity distributions of the brightest, non-variable
point sources should provide these calibration values.  This can, however, 
only be applied to sources for which the positions on the sky are well 
determined in both coordinates through the positional sphere reconstruction. 
For the {\it Planck} mission this condition limits its application to point 
sources near to the ecliptic poles.

Additional information can be obtained from transits of planets and minor 
planets, for which the absolute coordinates at any time during the mission 
are known to a much higher accuracy than required for the {\it Planck} 
calibrations, and which have the advantage of being visible to most or all 
detectors. To use these transits for opening angle calibrations requires,
however, accurate knowledge of the beam profile perpendicular to the scan 
direction as well as accurate predictions of their expected intensities, both
of which may be difficult to obtain.

\section[]{Iterations with calibrations and the ring analysis}
\label{Sec_iterate}

Iterations between the point-source catalogue and the ring reductions
are necessary to obtain a properly calibrated geometric reference 
system for the observations. Without these calibrations in place 
interpretation of the data in the form of (partial) maps will be of limited 
value, especially towards the higher frequencies in the power spectrum.

Iteration with the BPSC is also required to assign ordinates to point-source 
transits, which is essential in calibrating the two-dimensional 
beam profile. Using the BPSC for the identification of point sources in
the final ring reductions ensures that the background signals for all all
rings contain compatible information. Inconsistent point-source subtraction
would lead to harmonic signals in the ring analysis that can not be combined
in the harmonic map-making.  

An iteration with the harmonic map-making \cite{C1} for the half-year data is  
required to remove the spectral shift in the data which results from the 
velocity vector of the satellite (see Section~\ref{sec_aberr}). While the 
CMB dipole affects only the CMB component in the frequency maps,  
the satellite's velocity vector affects all component on every frequency map 
in both aberration and Doppler shift.

Most of these iterations require complete reprocessing of the ring data
and are, as such, very time consuming. Without them, however, the scientific 
interpretation of the data will be subject to considerable uncertainty.  
  
\section{Conclusions}
\label{Sec_concl}

The harmonic data model, of which the first part was presented in the current 
paper, provides a high level of information preservation in the data 
reductions. It defines calibration requirements and methods and provides a 
clear path from observed quantities to the scientific products. The latter
is important, as the interpretation of the science products requires  reliable 
knowledge of their statistical characteristics, which are well defined in
the harmonic model.

Although the methods presented in the current paper were developed as part of 
the harmonic data model, most would also be useful when using the more 
traditional pixelisation methods. This applies, for example, to the iterative
cycle of the ring reductions, BPSC construction and the geometric calibrations.
Phase binning of the data can also be used with pixel-based methods, as it 
provides a means for short-term response 
calibrations, point source recognition and data compression.

Further work is in progress in areas of point-source recognition from the ring 
data, and the cross-identification of point sources as detected on different 
rings.

The part of the {\it Planck} data processing presented in this paper will be
very demanding. An estimated half a million rings will be produced by the HFI
per year of observations. The calibration requirements will make it 
necessary to process each ring at least three times to get all geometric and 
beam profile calibrations implemented. The processing of such large quantities
of data requires careful planning.

\section*{Acknowledgments}

This work benefited from useful discussions with several members of
the \emph{Planck} collaboration, in particular we thank Neil Turok,
Martin Bucher and Rob Crittenden.
ADC acknowledges a PPARC Postdoctoral Research Fellowship.
DJM, MAJA, DM and VS are supported by PPARC under grant RG28936.

\appendix

\section[]{Planck attitude dynamics}
\label{App_att}

The dynamics of the {\it Planck} satellite are largely determined by the following 
characteristics:
\begin{itemize}
\item The total mass and its variation over the mission;
\item The inertia tensor and its variation over the mission;
\item The position of the centre of gravity (CoG);
\item The actual scan velocity, which will be very close to a fixed nominal 
value;
\item The size, position and optical characteristics of the solar panel.
\end{itemize}

\subsection[]{Input parameters}
\label{sec_input}

Early models by Matra Marconi (Dynamics and Pointing, 260/CDA/NT/83.95) have 
provided an initial set of values for the satellite characteristics, which we 
can use to derive a model for the satellite attitude. Although changes in 
these parameters can be expected for the final realisation of the {\it Planck}
satellite, the general character of the results presented here will not be 
much affected.

We use the following values:
\begin{itemize}
\item The total mass is 772~kg;
\item The inertia tensor (defined in the SRS coordinate system, see the next 
section) is given by 
\[{\mathbfss I} = \left[\matrix{699 & 4.0 & 4.5 
\cr 4.0 & 766 & 4.2 \cr 4.5 & 4.2 & 970}\right] \ {\mathrm kg~m^2},\]
where the actual values used are only of relevance in as far as that they
reproduce the approximate characteristics of the tensor;
\item The position of the CoG, in the same reference system as defined above, 
is given by 
\[\left[\matrix{x_0\cr y_0\cr z_0}\right] = \left[\matrix{0.0319\cr 
-0.0315\cr \-0.7282}\right] \ \ {\mathrm m};\]
\item The diameter of the solar panel is 4.5~m, with a secular reflection 
coefficient $C_{\mathrm s}=0.17$, and a diffuse reflection coefficient $C_{\mathrm d}=0.10$;
\item The solar radiation pressure is $4.5\times10^{-6}$Nm$^{-2}$.
\end{itemize}

The solar panel shields the rest of the satellite from solar radiation. Its 
position, size and optical characteristics therefore determine the solar 
radiation force experienced by the satellite. The position of the CoG then 
determines how much of this force is experienced as a torque (see for example
Spence, 1978). 

\subsection[]{The inertia tensor and coordinate alignments}
\label{sec_inert}

The inertia tensor given above is not diagonal, i.e.\ the principal axes of
the satellite do not align with the principal axes of the inertia tensor. We 
define two reference systems; the Satellite Reference 
System (SRS), aligned with the principle axes of the satellite, and the 
Inertia Reference System (IRS), aligned with the principle $z$-axis of the 
inertia tensor. The origin of both systems is the CoG.

The SRS is defined as follows. The $z$-axis is normal to the solar panel, goes
through the CoG and is pointing away from the Sun. The $x$-axis lies in the 
plane through the $z$-axis and the vector from the CoG to the FRP as projected 
onto the sky, and is 
orthogonal to the $z$-axis. The $y$-axis completes the right-handed triad.

The IRS is defined such that the off-diagonal elements $I_{xz}$ and 
$I_{yz}$ of the inertia tensor are zero. This is obtained by two small 
rotations $\alpha_{1,2}$. A third rotation ($\alpha_3$) is added to make
$z'$, $x'$ and the direction of the FRP on the sky coplanar.
The three rotations are clockwise around the $x$, $y$ and $z$ axes 
respectively, and define a rotation matrix
\begin{equation}
{\mathbfss R} \approx \left[\matrix{1 & -\alpha_3 & \alpha_2 \cr \alpha_3 & 1 & 
-\alpha_1 \cr -\alpha_2 & \alpha_1 & 1}\right].
\end{equation}
The inertia tensor ${\mathbfss I}'$ in the IRS is related to the 
inertia tensor ${\mathbfss I}$ in the SRS through
\begin{equation}
{\mathbfss I}' = {\mathbfss R}{\mathbfss I}{\mathbfss R}^T.
\end{equation}
The physical meaning of the angles $\alpha_{1,2,3}$ is simple: 
$\alpha_1$ is the rotation of the focal plane relative to the ring;
$\alpha_2$ is the difference between the nominal and actual opening
angle for the centre of the focal plane; $\alpha_3$ is an overall phase shift. 
 
Due to the depletion of consumables, the values in the inertia tensor cannot 
be assumed constant. The angles $\alpha_{1,2,3}$ will change over the 
mission as a result of this, $\alpha_1$ and $\alpha_2$ will require 
calibration during the mission, while $\alpha_3$ corrections are absorbed
in the zero-phase corrections that have to be applied to each circle. 

\subsection[]{The attitude dynamics}
\label{sec_att_dyn}

The attitude dynamics of a satellite describe the motion of the satellite
around its CoG in a suitably defined reference system. We consider the
satellite to be a rigid body, but the effects of deviations from this
assumption still need to be investigated: a possible source of violation of 
the rigidity assumption is the circulation of cooler liquids. As a rigid
body, the satellite's motions are described by the Euler equation, defined 
here in the IRS:
\begin{equation}
{\mathbfss I}'\frac{{\mathrm d}\bomega}{{\mathrm d}t} = \bmath{N}' - 
\bomega\times{\mathbfss I}'\bomega,
\label{equ_eul}
\end{equation}
where ${\bmath N^\prime}$ represents the external and internal torques acting on
the satellite, and $\bomega$ the inertial rates around the axes of the IRS.
Of the three components of $\bomega$, $\omega_z$ is by far the dominant one,
with a nominal value of 0.1047~rad~s$^{-1}$. As a result, the cross product
in equation~(\ref{equ_eul}) is only relevant for the $x$ and $y$ 
coordinates. This allows us to approximate equation~(\ref{equ_eul}) as
\begin{eqnarray}
\frac{{\mathrm d}\omega_x}{{\mathrm d}t} &\approx& N^\prime_x/{I^\prime_{xx}} - 
f_1\omega_y\omega_z + \frac{I^\prime_{xy}}{I^\prime_{xx}}\omega_x\omega_z, 
\nonumber\\
\frac{{\mathrm d}\omega_y}{{\mathrm d}t} &\approx& N^\prime_y/{I^\prime_{yy}} + 
f_2\omega_x\omega_z - \frac{I^\prime_{xy}}{I^\prime_{yy}}\omega_y\omega_z, 
\nonumber\\
\frac{{\mathrm d}\omega_z}{{\mathrm d} t} &\approx& N^\prime_z/{I^\prime_{zz}}, \label{eq_appr}
\end{eqnarray}
where the third terms in the first two of these equations are ignored in the 
analytic model, and where
\begin{eqnarray}
f_1 = {{I^\prime_{zz}}-{I^\prime_{yy}}\over {I^\prime_{xx}}}, \nonumber \\
f_2 = {{I^\prime_{zz}}-{I^\prime_{xx}}\over {I^\prime_{yy}}}.
\end{eqnarray}
The quantity $\lambda=1+\sqrt{f_1f_2}$ is generally used in specifying the 
dynamic characteristics of the inertia tensor. In the preliminary 
specifications for the {\it Planck} satellite $\lambda\approx1.35$. Here we 
use instead $\gamma=\lambda-1$ in the development of the dynamic equations.

\subsection[]{The solar radiation torque}

The solar radiation acts only on the circular solar panel, which is shielding
the satellite. Relative to the Sun, the satellite's orientation is 
described by the solar aspect angle $\xi$ between the spin axis and the 
direction of the Sun, and the spin phase $\psi$, measured in the direction 
of rotation from the transit of the focal plane through the great circle
defined by the directions of the Sun and the spin axis 
(see Fig.~\ref{fig_ang}). This defines the unit vectors $\hat{\bmath s}'$ in 
the IRS and $\hat{\bmath s} = {\mathbfss R}^{-1}\hat{\bmath s}'$ in the SRS, 
from the solar panel in the direction of the Sun as 
\begin{equation}
\hat{\bmath{s}'}=\left[\matrix{-\cos\psi\sin\xi\cr 
\sin\psi\sin\xi\cr -\cos\xi}\right], \qquad
\hat{\bmath{s}}\approx\left[\matrix{-\cos\psi\sin\xi + \alpha_2\cr 
\sin\psi\sin\xi - \alpha_1\cr -\cos\xi}\right],
\label{eq_s1}
\end{equation}
where $\xi$ is assumed to be very small ($\le~0.03$~rad.).
Following Spence (1978), the solar radiation force $\bmath{F}_{\mathrm rad}$ 
on a surface area ${\mathrm d} A$ due to reflection and absorption at the 
solar panel is given by
\begin{eqnarray}
{\mathrm d}\bmath{F}_{\mathrm rad} &=& -P\left[(1-C_{\mathrm s})\hat{\bmath s}+2(C_{\mathrm s}\cos\xi+
{1\over 3}C_{\mathrm d})\hat{\bmath n}\right]\cos\xi{\mathrm d}A,\nonumber \\
\label{eq_force}
\end{eqnarray} 
where $C_{\mathrm s}$ and $C_{\mathrm d}$ are defined in 
Section~\ref{sec_input}, and $\hat{\bmath n}$ is the unit vector normal to the 
solar panel [$\hat{\bmath n} = (0,0,-1)$]. The solar radiation pressure $P$ 
is $4.5\times10^{-6}$~Nm$^{-2}$. We introduce the coefficients $G$ and $H$:
\[G=(1-C_{\mathrm s}), \qquad H=(1+C_{\mathrm s})\cos\xi + {2\over 3}C_{\mathrm d}, \]
with which equation~(\ref{eq_force}) can be written as
\begin{eqnarray}
{\mathrm d}\bmath{F}_{\mathrm rad} &=& -P\left[\matrix{
G(-\cos\psi\sin\xi+\alpha_2)\cr G(\sin\psi\sin\xi-\alpha_1)\cr -H}
\right]\cos\xi{\mathrm d}A.\nonumber \\
\label{eq_force_c}
\end{eqnarray} 
The solar radiation torque acting on the satellite is obtained by
integrating the outer product of the force and the separation of the surface
element from the CoG over the surface area of the solar panel. If $(\rho,\phi)$
are polar coordinates defined relative to the centre of the solar panel, then
the position vector ${\bmath r}$ of the surface element ${\mathrm d}A$ is 
given by
\begin{equation}
{\bmath r} = \left[\matrix{\rho\cos\phi + x_0\cr \rho\sin\phi + 
y_0\cr z_0}\right],
\label{eq_r}
\end{equation}
where $(x_0,y_0,z_0)$ is the position of the centre of the solar panel 
relative to the CoG, measured in the SRS. The torque acting on the satellite 
is now given by
\begin{equation}
{\bmath N} = \int\int {\bmath r}\times{\mathrm d}{\bmath F}_{\mathrm rad},
\label{eq_torq}
\end{equation}
and the surface element by
\begin{equation}
{\mathrm d}A = \rho{\mathrm d}\rho{\mathrm d}\phi.
\label{eq_surf}
\end{equation}
In the integration, the terms containing $\phi$ in equation~(\ref{eq_r}) 
cancel, leaving only the effects of the offset of the CoG:
\begin{equation}
{\bmath N} \approx {\bmath N}_0 + {1\over 2} G
\sin 2\xi\left[\matrix{ 0 & z_0 \cr z_0 & 0 \cr 
 -y_0 & -x_0}\right]\left[\matrix{ \cos\psi\cr \sin\psi}\right]{\cal F},
\label{eq_torq2}
\end{equation}
where non-modulated part of the torque, ${\bmath N}_0$, is given by
\begin{equation}
{\bmath N}_0 = \cos\xi\left[
\matrix{y_0H-z_0G\alpha_1\cr -x_0H-z_0G\alpha_2\cr
G(x_0\alpha_1+y_0\alpha_2)}\right]{\cal F},
\label{eq_N0}
\end{equation}
and the total force ${\cal F}=7.16\times10^{-5}$~N. 
In the ISR the torques are given by
\begin{equation}
{\bmath N}^\prime_0 = \cos\xi\left[
\matrix{y_0H-z_0G\alpha_1 - x_0H\alpha_2\cr -x_0H-z_0G\alpha_2+y_0H\alpha_3\cr
(G-H)(x_0\alpha_1+y_0\alpha_2)}\right]{\cal F},
\label{eq_N0prime}
\end{equation}
while the product of $\sin2\xi$ and $\alpha_{1,2,3}$ are small enough to 
leave the second part in equation~(\ref{eq_torq2}) unchanged.

The inertia tensor given in Section~\ref{sec_input} defines an order of 
magnitude for the off-diagonal elements, based on the Matra Marconi document, 
which provides estimates of 0.03 rad for $\alpha_1$ and $\alpha_2$. The typical
resulting solar radiation 
torques are at a level of $2\times10^{-6}$~Nm in $x$ and $y$, and 10$^{-7}$ in 
$z$, equivalent to accelerations at a level of 1 to 0.1~milliarcsec~s$^{-2}$. 
Scanning at a tilt of 10\degr\ adds a modulated torque with an amplitude of 
about $7.5\times10^{-6}$~Nm in $x$ and $y$, and $3.8\times10^{-7}$~Nm in $z$. 
In all cases, the effect of the solar radiation is small with respect to the 
beam sizes of the {\it Planck} detectors, and the only relevant contribution 
is the acceleration or deceleration of the spin rate, which will accumulate to 
give barely significant phase shifts over a one hour scan period.

\subsection[]{Analytic model for the inertial rates}

In its slightly simplified form, given by the first two terms only, 
equations~(\ref{eq_appr}) can be solved to provide a model for the evolution 
of the inertial rates around the satellite axes. Assuming $\omega_z t = 
\psi$, we find
\begin{eqnarray}
\omega_x &=& -{N^\prime_{y,0}\over I^\prime_{yy}f_2\bar{\omega}_z} -b_x\cos\bar{\omega}_zt 
\nonumber \\
 & &  +\omega_1\cos(\bar{\omega}_z\gamma t) - 
\omega_2\sin(\bar{\omega}_z\gamma t), \nonumber \\
\omega_y &=& {N^\prime_{x,0}\over I^\prime_{xx}f_1\bar{\omega}_z}+b_y\sin\bar{\omega}_zt
\nonumber \\
& & +{\gamma\over f_1}\Bigl[\omega_2\cos(\bar{\omega}_z\gamma t) + 
\omega_1\sin(\bar{\omega}_z\gamma t)\Bigr],\nonumber \\
\omega_z &=& \omega_3+{N^\prime_{z,0}\over I^\prime_{zz}}t-{G\sin2\xi\over2\bar{\omega}_z
I^\prime_{zz}}\bigl(y_0\sin\bar{\omega}_zt - x_0\cos\bar{\omega}_zt\bigr){\cal F},
\nonumber \\
\label{eq_xy}
\end{eqnarray}
where $N^\prime_{x,0}$, $N^\prime_{y,0}$ and $N^\prime_{z,0}$ are the components of $\bmath{N}_0$
as defined in equation~(\ref{eq_N0}), and $\bar{\omega}_z$ is the nominal scan 
velocity, and $(\omega_1,\omega_2,\omega_3)$ are integration constants. Also
\begin{eqnarray}
b_x&=&{1\over 2}{z_0G\sin 2\xi(2I^\prime_{yy}/I^\prime_{zz}-1)\over(I^\prime_{xx}+I^\prime_{yy}-I^\prime_{zz})
\bar{\omega}_z}~{\cal F},\nonumber \\
b_y&=&{1\over 2}{z_0G\sin 2\xi(2I^\prime_{xx}/I^\prime_{zz}-1)\over(I^\prime_{xx}+I^\prime_{yy}-I^\prime_{zz})
\bar{\omega}_z}~{\cal F},
\label{eq:bxby}
\end{eqnarray}
where $F$, as defined in equation~(\ref{eq_N0prime}), is the total solar radiation 
force, and $\bar{\omega}_z\gamma$ is the nutation frequency, equivalent to a 
nutation period of just under 3~minutes for $\gamma=0.35$ and $\omega_z$ equal
to 1 rpm. The satellite will undergo nutation damping after each repositioning 
of the spin axis. This reduces the rotation rates around the $x$ and $y$ 
axes to a level low enough not to cause significant disturbances on the 
measurements. If we assumed that, as a result of the nutation damping, the 
rotation rates at the start of a scan are equal to 
$(\omega_{x,0},\omega_{y,0},\omega_{z,0})$, then 
$(\omega_1,\omega_2,\omega_3)$ are given by
\begin{eqnarray}
\omega_1 &=& \omega_{x,0}+ \frac{N^\prime_{y,0}}{I^\prime_{yy}\bar{\omega}_z f_2} + 
b_x,\nonumber\\
\omega_2 &=&  \frac{f_1}{\gamma}\Bigl[\omega_{y,0}-\frac{N^\prime_{x,0}}
{I^\prime_{xx}\bar{\omega}_z f_1}\Bigr],\nonumber \\
\omega_3 &=& \omega_{z,0}- x_0\frac{G\sin2\xi}{2\bar{\omega}_z I^\prime_{zz}}{\cal F},
\label{eq_om_init}
\end{eqnarray}
where the reference time $t=0$ is the first instance of $\psi=0$ after the 
end of the nutation damping process. 

\subsection[]{Analytic model for the error angles}

The differences between the nominal and actual orientations of the IRS
axes are described by a set of angles generally referred to as the
Tait-Bryan angles $(T_z,T_y,T_x)$. The relations between increments in the 
Tait-Bryan angles and the inertial rates are described by the equations give
below, where it is assumed that the inertial rates resulting from the 
nominal scanning law are given by $\bar{\omega}_z$, and the actual
inertial rates by $(\omega_x,\omega_y,\omega_z)$.

The inertial rate for the nominal scanning law needs to be transformed to 
inertial rates around the actual satellite axes:
\begin{equation}
\left[\matrix{\omega_{x,i}\cr\omega_{y,i}\cr\omega_{z,i}}\right] = 
\bar{\omega}_z
\left[\matrix{\sin T_x\sin T_z - \cos T_x\cos T_z\sin T_y\cr
\sin T_x\cos T_z + \cos T_x\sin T_z\sin T_y\cr \cos T_x\cos T_y\hfill }\right].
\end{equation}
The differences between the actual rates and the projected nominal rates 
are transformed to corrections to the Tait-Bryan angles:
\begin{equation}
\dot{\bmath T} = \left[\matrix{
\cos T_z/\cos T_y &  -\sin T_z/\cos T_y & 0\cr \sin T_z &  \cos T_z & 0 \cr  
- \cos T_z\tan T_y & \sin T_z\tan T_y & 1 }\right]\delta\bomega,
\end{equation}
where
\begin{equation}
\delta\bomega = \left[\matrix{\omega_x-\omega_{x,i} \cr \omega_y-
\omega_{y,i} \cr \omega_z-\omega_{z,i}}\right].
\end{equation} 
The first set of equations can be approximated by
\begin{equation}
\left[\matrix{\omega_{x,i}\cr\omega_{y,i}\cr\omega_{z,i}}\right] \approx
\bar{\omega}_z 
\left[\matrix{-T_y\cr T_x \cr 1 }\right].
\end{equation}
Similarly, the second set of equations becomes
\begin{eqnarray}
\dot T_x \approx & (\omega_x+T_y\bar{\omega}_z),\nonumber\\ 
\dot T_y \approx & (\omega_y-T_x\bar{\omega}_z),\nonumber\\
\dot T_z \approx & (\omega_z-\bar{\omega}_z),  
\label{eq_difft}
\end{eqnarray}
where the relations for $(\omega_x,\omega_y)$ given in equation~(\ref{eq_xy})
can be used to obtain an approximate solution for $(T_x,T_y)$. 
Equation~(\ref{eq_xy}) is written as
\begin{eqnarray}
\omega_x &=& -a_x - b_x\cos\bar{\omega}_zt + \omega_1\cos(\bar{\omega}_z\gamma t) - \omega_2\sin(\bar{\omega}_z\gamma t), \nonumber \\
\omega_y &=& a_y + b_y\sin\bar{\omega}_zt  \nonumber \\
&&+{\gamma\over f_1}\bigl[\omega_2\cos(\bar{\omega}_z\gamma t) + \omega_1\sin(\bar{\omega}_z\gamma t)\bigr].
\label{eq_xy2}
\end{eqnarray}
Similarly, the angles $(T_x,T_y)$ are represented by
\begin{eqnarray}
T_x &=& A_x + B_x\cos\bar{\omega}_zt + C_x\sin\bar{\omega}_zt + tF\cos\bar{\omega}_zt \nonumber \\
&&+ D_x\cos\bar{\omega}_z\gamma t + E_x\sin\bar{\omega}_z\gamma t, \nonumber \\
T_y &=& A_y + B_y\cos\bar{\omega}_zt - C_y\sin\bar{\omega}_zt - tF\sin\bar{\omega}_zt \nonumber \\ 
&&+ D_y\cos\bar{\omega}_z\gamma t + E_y\sin\bar{\omega}_z\gamma t.
\label{eq_txty}
\end{eqnarray}
\begin{figure*}
\centerline{\psfig{file=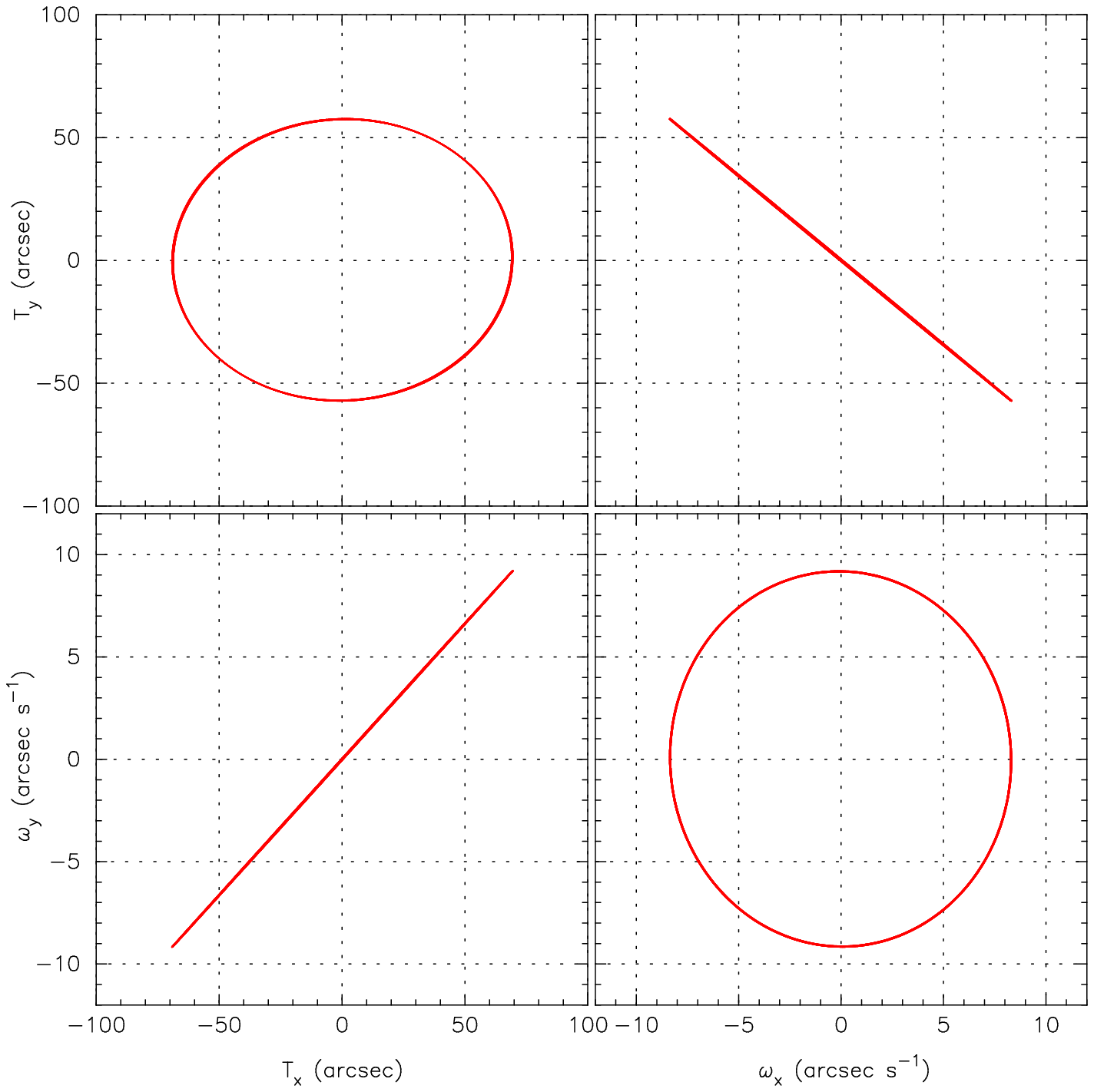,height=11truecm}}
\caption{A typical evolution pattern for inertial rates and error angles for 
$\xi=10$~degrees. The two graphs in the upper right and lower left corners 
show the almost perfect relations between rates and error angles.}
\label{fig_1a}
\end{figure*}
Substituting equations~(\ref{eq_xy2}) and (\ref{eq_txty}) in 
equation~(\ref{eq_difft}), we find
\begin{eqnarray}
A_x = a_y/\bar{\omega}_z&, & A_y = a_x/\bar{\omega}_z, \nonumber\\
C_x = C_1&, & B_y = {(b_x-b_y)\over2\bar{\omega}_z} + C_1, \nonumber\\
B_x = C_2&, & C_y = C_2, \nonumber\\
D_x = \frac{\omega_2(\gamma/f_1-\gamma)}{\bar{\omega}_z(1-\gamma^2)}&, & 
D_y = \frac{\omega_1(f_2-1)}{\bar{\omega}_z(1-\gamma^2)},\nonumber\\
E_x = \frac{\omega_1(\gamma/f_1-\gamma)}{\bar{\omega}_z(1-\gamma^2)}&, &
E_y = -\frac{\omega_2(f_2-1)}{\bar{\omega}_z(1-\gamma^2)},\nonumber\\
F = \frac{-(b_x+b_y)}{2},&& 
\end{eqnarray} 
where $C_1$ is a constant, depending on the choice of the reference 
position of the satellite's $z$-axis. If $(T_{x,0},T_{y,0})$ are the 
Tait-Bryan angles at time zero, then for $\xi=0$ the following 
relations are obtained:
\begin{eqnarray}
T_{x,0} &=& {a_y\over\bar{\omega}_z}+C_1 +\frac{\omega_2(\gamma/f_1-\gamma)}
{\bar{\omega}_z(1-\gamma^2)}, \nonumber \\
T_{y,0} &=& {a_x\over\bar{\omega}_z}+C_2 +\frac{\omega_1(f_2-1)}
{\bar{\omega}_z(1-\gamma^2)},
\label{eq_tzero}
\end{eqnarray}
where $a_x$ and $a_y$ are relatively small with respect to the last terms. 
It is possible to choose the reference position for the Tait-Bryan angles
such that $C_1$ and $C_2$ are equal to zero. This reduces the description
of the motion of the satellite $z$-axis to a simple ellipse in the 
$(T_x,T_y)$ plane.

The coefficient for $F$ is a potential source of instability. An evaluation of
the expected value of $F$, using equation~(\ref{eq:bxby}), gives 
\begin{equation}
F = \frac{-z_0G\sin2\xi}{2I^\prime_{zz}\bar{\omega}_z}{\cal F}.
\end{equation}
Using the input parameters specified in Section~\ref{sec_input} shows that 
the effect is 
quite small. From the values given we estimate a maximum drift of 
$166\times\sin 2\xi$~arcsec over a 1~hour period. With $\xi$ having a maximum 
value of 10~degrees, this is noticeable at the resolutions used by the {\it 
Planck} instruments.

It can be verified that for $\xi=0$ and $B=C_1=C_2=0$, and a fixed value of 
$\bar{\omega}_z$, linear relations exist between 
$T_x$ and $\omega_y$ and between $T_y$ and $\omega_x$. The ratios are
\begin{eqnarray}
R_1\equiv&{T_x-a_y/\bar{\omega}_z\over\omega_y} =& \frac{1}{\bar{\omega}_z}\frac{1-f_1}{1-\gamma^2},
\nonumber\\
R_2\equiv&{T_y-a_x/\bar{\omega}_z\over\omega_x} =& \frac{-1}{\bar{\omega}_z}\frac{1-f_2}{1-\gamma^2}.
\label{eq_r1_r2}
\end{eqnarray}
These relations, shown in Fig.~\ref{fig_1a}, get only slightly disturbed by 
non-zero values for the 
solar aspect angle. They show that the satellite dynamics is dominated by
the effects of the residual velocities around the $x$ and $y$ axes.

\subsection[]{Numerical integration of the attitude model}

Numerical integration of the Euler equation and the resulting rates
can be performed without any of the approximations made in the analytic model;
this provides a useful cross check on the analytic results.
The integration over the Euler equation requires a set of inertial rates at
the starting time. These have been chosen such that they represent the nominal 
rates: zero in $\omega_x$ and $\omega_y$, 6~degrees per second in $\bar{\omega}_z$. 
At time-steps of 0.01~s the torque acting on the satellite is 
developed, as well as the crossproducts of the rates 
${\mathbfss I}^{-1}\omega\times{\mathbfss I}\omega$. This integration 
provides a record of the inertial rate development.

Similarly, the integration over the Tait-Bryan angles requires an assumption 
about the starting values. For $T_x$ and $T_y$ equation~(\ref{eq_tzero}) is 
used to calculate the starting values, with $C_1=C_2=0$, while the $T_z$ 
starting value is set to zero.

\subsubsection[]{Comparison with the analytic results}

\begin{figure*}
\centerline{\psfig{file=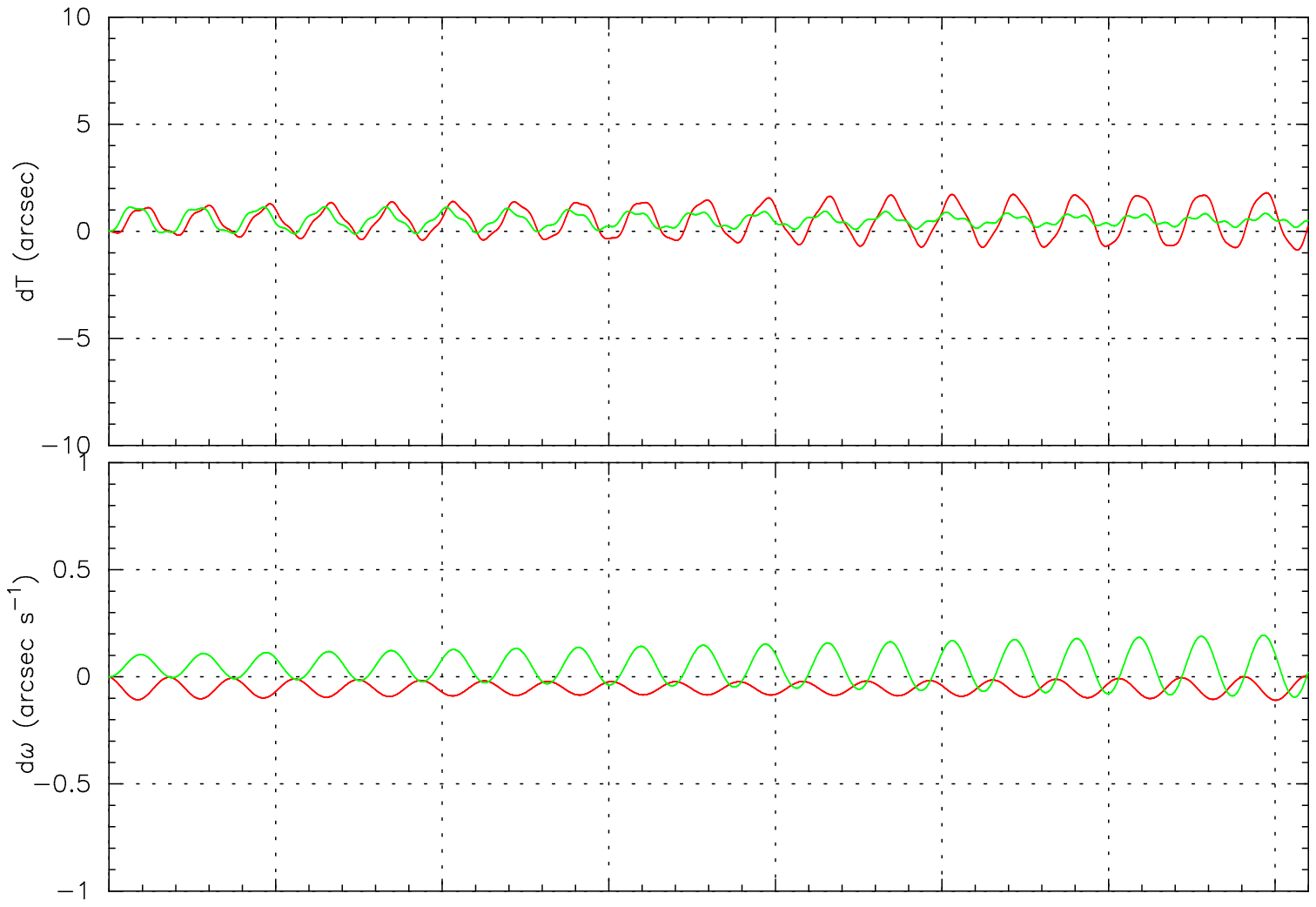,width=12truecm}}
\centerline{\psfig{file=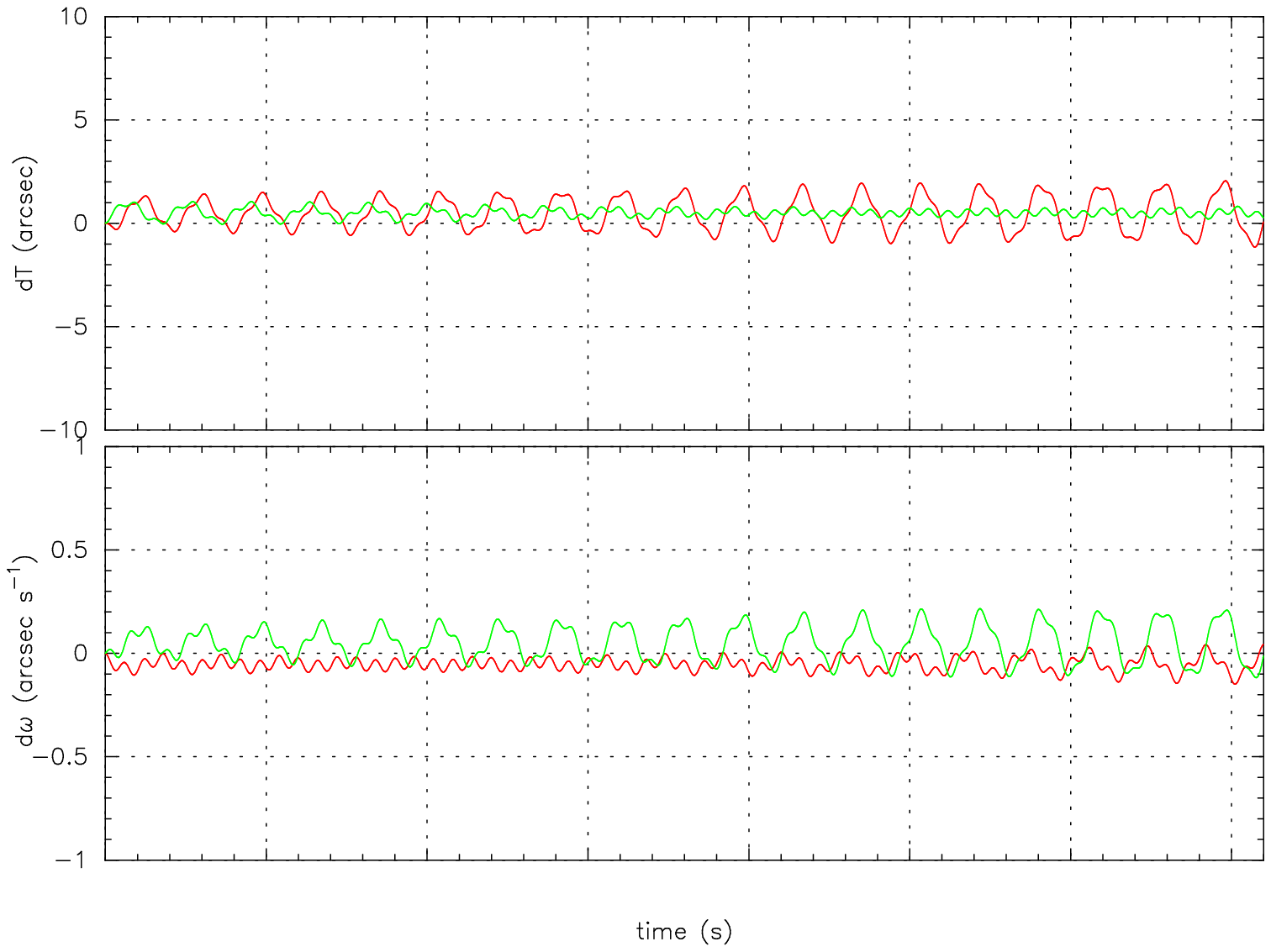,width=12truecm}}
\caption{The differences between rates and error angles around the $x$ and 
$y$ axes as obtained from analytic modelling and from numerical integration. 
Top two graphs: a one-hour scan with $\xi=0$; bottom two graphs: 
$\xi=10$~degrees. The two curves in each graph show the variations for the 
$x$ and $y$ axes.}
\label{fig_2}
\end{figure*}
The results of the analytic approximation are compared with the numerical 
integrations in Fig.~\ref{fig_2}. These comparisons show that over a one-hour 
period the analytic approximation adds noise at a level of a few arcsec 
to the more accurate numerical integration. The main component of the noise
originates from ignoring the $I^\prime_{xy}$ term in the development of
equation~(\ref{eq_appr}). Including this term affects the amplitudes $\omega_1$
and $\omega_2$, and adds to these a small time-dependent term, which can,
if necessary, also be estimated analytically. However, these corrections
appear to be insignificant for {\it Planck}. The positional error in the 
analytic model is of the order of a few arcsec, which again is 
insignificant for {\it Planck}.

Given this level of agreement between the analytic and numerical models, 
it will be sufficient for the {\it Planck} simulations to use the analytic
model, and to simulate only the initial conditions for a scan. In the same
way, the conditions during re-positioning of the $z$-axis can be simulated.
The simulations, as well as the analytic model, show that a solar aspect angle 
of $10^\circ$ or less has very little influence on the satellite's attitude, 
unless this is examined at accuracies much higher than those required for 
{\it Planck}.

\section[]{Projections on the reference circle}
\label{app_wobble}

In this appendix we describe the transformations for positions and rates
from a system in which the spin axis describes a nutation ellipse in the
satellite reference frame to a reference circle with its axis at the centre
of this ellipse. 

For $\xi=0^\circ$, the nutation ellipse is described in the satellite 
reference system by
\begin{eqnarray}
T_x &\approx& A_x + D_x\cos\bar{\omega}_z\gamma t + E_x\sin\bar{\omega}_z\gamma t \nonumber \\
T_y &\approx& A_y + D_y\cos\bar{\omega}_z\gamma t + E_y\sin\bar{\omega}_z\gamma t,
\end{eqnarray}
with the coefficients as defined in equation~(\ref{eq_txty}). In 
equation~(\ref{eq_om_init})
it can in addition be assumed that $(\omega_{x,0},\omega_{y,0})$ are by far 
the dominant coefficients in $(\omega_1,\omega_2)$. This also implies that
the coefficients $(A_x,A_y)$ will be relatively small with respect to
$(\omega_1,\omega_2)$, so that we are left with 
\begin{eqnarray}
\omega_x &\approx& \omega_{x,0}\cos(\bar{\omega}_z\gamma t) - 
\frac{f_1}{\gamma}\omega_{y,0}\sin(\bar{\omega}_z\gamma t), \nonumber \\
\omega_y &\approx&\omega_{y,0}\cos(\bar{\omega}_z\gamma t) + {\gamma\over f_1}
\omega_{x,0}\sin(\bar{\omega}_z\gamma t), \nonumber \\
T_x &\approx& R_1\bigl[\omega_{y,0}\cos(\bar{\omega}_z\gamma t) + 
{\gamma\over f_1}\omega_{x,0}\sin(\bar{\omega}_z\gamma t)\bigr], \nonumber \\
T_y &\approx& R_2\bigl[\omega_{x,0}\cos(\bar{\omega}_z\gamma t) - 
{f_1\over \gamma}\omega_{y,0}\sin(\bar{\omega}_z\gamma t)\bigr],
\label{eq_om_t}
\end{eqnarray}
where $R_1$ and $R_2$ are functions of $f_1$, $f_2$ and $\bar{\omega}_z$, as defined
in equation~(\ref{eq_r1_r2}), and $\gamma=\sqrt{f_1f_2}$ as defined in 
Section~\ref{sec_att_dyn}.
By choosing a suitable reference time $t_0$ equation~(\ref{eq_om_t}) can be
written as 
\begin{eqnarray}
\omega_x &\approx& \omega_{x,0}\cos\bigl[\bar{\omega}_z\gamma (t-t_0)\bigr], 
\nonumber \\
\omega_y &\approx&{\gamma\over f_1}\omega_{x,0}\sin\bigl[\bar{\omega}_z\gamma 
(t-t_0)\bigr], \nonumber \\
T_x &\approx& R_1{\gamma\over f_1}\omega_{x,0}\sin\bigl[\bar{\omega}_z\gamma
(t-t_0)\bigr], 
\nonumber \\
T_y &\approx& R_2\omega_{x,0}\cos\bigl[\bar{\omega}_z\gamma(t-t_0)\bigr],
\label{eq_om_t0}
\end{eqnarray}
The mean spin axis is assumed to be a fixed point in space, in the centre of 
the ellipse described by $(T_x,T_y)$, and the spin velocity is assumed to be 
the nominal spin velocity, $\omega_z = \bar{\omega}_z$. The offset angles can be
defined with respect to a fixed reference system relative to the mean spin
axis:
\begin{eqnarray}
T_x' &=& T_x\cos\psi - T_y\sin\psi,\nonumber \\
T_y' &=& T_x\sin\psi + T_y\cos\psi.
\label{eq_T_tran}
\end{eqnarray}
The angle $\psi$ is measured on the ring from its crossing of the 
reference circle (see Fig.~\ref{fig_ang}). 
A point on the ring is given by the direction cosines
\begin{equation}
O_i=\left[\matrix{\cos\psi\sin\alpha\cr \sin\psi\sin\alpha\cr 
\cos\alpha}\right],
\label{eq_dircos}
\end{equation}
where $\alpha$ is the opening angle for the reference circle. The offset
angles $(T_x',T_y')$ create small offsets in the actual angles $(\psi,\alpha)$
for a measurement. We ignore the very small rotation over $T_z$. The rotations
over $T_x'$ and $T_y'$ distort equation~(\ref{eq_dircos})
\begin{equation}
O_i=\left[\matrix{\cos\psi\sin\alpha + T_y'\cos\alpha\cr 
\sin\psi\sin\alpha - T_x'\cos\alpha \cr \cos\alpha +T_x'\sin\alpha\sin\psi-
T_y'\sin\alpha\sin\psi}\right],
\label{eq_dircosT}
\end{equation}
which can also be expressed as
\begin{equation}
O_i=\left[\matrix{\cos\psi\sin\alpha + \Delta\alpha\cos\alpha\cos\psi
-\Delta\psi\sin\alpha\sin\psi\cr 
\sin\psi\sin\alpha + \Delta\alpha\cos\alpha\sin\psi
+\Delta\psi\sin\alpha\cos\psi\cr \cos\alpha - \Delta\alpha\sin\alpha}\right].
\label{eq_dircosA}
\end{equation}
From these relations the expressions for 
$(\Delta\psi,\Delta\alpha)$ in terms of $(T_x',T_y')$ follow:
\begin{eqnarray}
\Delta\psi &=& (-T_x'\cos\psi - T_y'\sin\psi)/\tan\alpha,\nonumber \\
\Delta\alpha &=& -T_x'\sin\psi + T_y'\cos\psi.
\end{eqnarray}
Substituting equation~(\ref{eq_T_tran}) we find
\begin{eqnarray}
\Delta\psi &=& -T_x/\tan\alpha,\nonumber \\
\Delta\alpha &=& + T_y.
\end{eqnarray}
With an opening angle $\alpha$ of 80 to 85 degrees, the effect of the 
spin-axis wobble on the scan phases is very small. The maximum amplitude
is expected to be less than 0.12~arcmin. This may be visible, though, in the
star mapper data. The effect of $T_x$ on the orientation of the focal plane
is even smaller.

The $\alpha$ variations can be relevant for the high-frequency detectors.
A change in position of the beam by $\pm0.75$~arcmin for a point source close 
to the steepest slope of the beam will create significant variations in the 
signal over a TOP. If the wobble parameters can be reconstructed from 
the star mapper data, then some of this variation may be accounted for in the 
reductions. 

\bsp

\label{lastpage}

\end{document}